\documentclass[preprint,tightenlines,showpacs,showkeys,floatfix,nofootinbib,
superscriptaddress,fleqn]{revtex4}
\usepackage{graphicx}
\usepackage{bm}
\usepackage{dcolumn}
\usepackage{amsmath}
\usepackage{amsfonts}
\usepackage{amssymb}
\usepackage{color}

\begin{document}

\preprint{INHA-NTG-09/2008}
\title{Axial-vector transitions and strong decays of the baryon
antidecuplet in the self-consistent SU(3) chiral quark-soliton model}
\author{Tim Ledwig}
\email{Tim.Ledwig@tp2.rub.de}
\affiliation{Institut f\"ur Theoretische Physik II, Ruhr-Universit\"
at Bochum, D--44780 Bochum, Germany}
\author{Hyun-Chul Kim}
\email{hchkim@inha.ac.kr} 
\affiliation{Department of Physics, Inha University, Incheon 402-751, 
Republic of Korea}
\author{Klaus Goeke}
\email{Klaus.Goeke@tp2.rub.de}
\affiliation{Institut f\"ur Theoretische Physik II, Ruhr-Universit\"
at Bochum, D--44780 Bochum, Germany}
\date{May 2008}

\begin{abstract}
We investigate the axial-vector transition constants of the
baryon antidecuplet to the octet and decuplet within the  
framework of the self-consistent SU(3) chiral quark-soliton
model.  Taking 
into account rotational $1/N_{c}$ and linear $m_{s}$ corrections and
using the symmetry-conserving quantization, we calculate the
axial-vector transition constants.  It is found that the 
leading-order contributions are generally almost canceled by the
rotational $1/N_{c}$ corrections.  Thus, the $m_{s}$ corrections turn
out to be essential contributions to the axial-vector constants.  The
decay width of the $\Theta^{+}\to NK$ transition is determined to be 
$\Gamma(\Theta\to NK)=0.71$ MeV, based on the result
of the axial-vector transition constant $g_A^* (\Theta \to
NK)=0.05$.  In addition, other strong decays of the baryon
antidecuplet are investigated. 
\end{abstract}

\pacs{12.39.Fe, 13.40.Em, 12.40.-y, 14.20.Dh}
\keywords{Axial-vector transition constants, chiral quark-soliton
model, pentaquark baryons, strong decays of SU(3) baryons} 

\maketitle
\section{Introduction}
Since the LEPS collaboration announced the evidence of the $\Theta^+$
existence~\cite{Nakano:2003qx}, which was motivated by
Ref.~\cite{Diakonov:1997mm} where its decay width was predicted to be
very small with its mass $1540$ MeV~\cite{Praszalowicz:2003ik} as
well, there has been a great deal of experimental and theoretical work
on the $\Theta^+$ (see, for example,
reviews~\cite{Hicks:2005gp,Goeke:2004ht} for the experimental and  
theoretical status before 2006).  However, a series of very recent
experiments conducted by the CLAS collaboration reported null
results of finding the $\Theta^+$~\cite{CLAS1,CLAS2,CLAS3,CLAS4} in
various reactions.  These CLAS experiments were dedicated ones with
high statistics.  The null results of the CLAS experiment imply that
the total cross sections for photoproductions of the $\Theta^+$ are
tiny.  In fact, the $95\%$ CL upper limits on the total production
cross sections for the $\Theta^+$ at 1540 MeV lie mostly 
in the range of of $0.3-0.8$ nb~\cite{CLAS1,CLAS2,CLAS4}.  In
Ref.~\cite{CLAS3} the $95\,\%$ CL upper limit on the $\gamma
d\to\Lambda\Theta^+$ total cross section is set to be 5 nb in the mass
range between 1.52 and $1.56\,\mathrm{GeV}/c^2$.  The KEK-PS E522
collaboration~\cite{Miwa:2006if} has performed the experiment 
searching for the $\Theta^+$ in the $\pi^- p \to K^- X$ reaction and
found a bump at $1530\,\mathrm{MeV}/c^2$ but with only
$(2.5-2.7)\,\sigma$ statistical significance.  The upper limit of
$\Theta^+$ production cross section in the $\pi^- p \to K^- \Theta^+$
reaction was extracted to be $3.9\mu\mathrm{b}$.  A later sequential
experiment at KEK, however, has observed no clear peak structure in
the $K^+p\to \pi^+ X$ reaction~\cite{Miwa:2007xk}, giving a $95\,\%$
CL upper limit of $3.5\,\mu\mathrm{b/sr}$ on the differential cross
section averaged over $2^\circ$ to $22^\circ$.    

In the meanwhile, the DIANA collaboration has continued to search for
the $\Theta^+$ in the $K^+n\to K^0 p$ reaction and has found a direct
formation of a narrow $pK^0$ peak with mass of $1537\pm
2\,\mathrm{MeV}/c^2$ and width of
$\Gamma=0.36\pm0.11\,\mathrm{MeV}$~\cite{DIANA2}.  Note that the  
former measurement by the DIANA collaboration for the $\Theta^+$ has
yielded the mass of the $\Theta^+$ $1539\pm2\,\mathrm{MeV}/c^2$ with
the decay width $\Gamma \le 9$ MeV~\cite{Barmin:2003vv}.  The SVD
experiment has also announced a narrow peak with the mass
$M= 1523\pm2(\mathrm{stat.})\pm3(\mathrm{syst.})\,\mathrm{MeV}/c^2$
in the inclusive reaction $pA\to pK_s^0+X$~\cite{SVD,SVD2008}.  The
LEPS collaboration also brings about positive new results indicating
the existence of the $\Theta^+$~\cite{LEPS2,Nakano}.  

Theoretically, it is of great importance to understand why the
$\Theta^+$ is rather elusive.  Actually, the cross sections of the
$\Theta^+$ photoproduction as well as of the mesonic production are
known to be very small.  One reason for this can be attributed to the
fact that the $K^*N\Theta^+$ coupling constant should be tiny, as was 
pointed out in Ref.~\cite{Miwa:2007xk}.  In fact,
Ref.~\cite{Kwee:2005dz} has shown that the tensor coupling constant 
for the $K^*N\Theta^+$ vertex is indeed very small.  In fact,
Ref.~\cite{Kwee:2005dz} has predicted even before the CLAS null
results that the cross section for the $\Theta^+$ photoproduction is
below the upper limits given by the above-mentioned CLAS 
experiments~\cite{CLAS1,CLAS2,CLAS4}.  Azimov et
al.~\cite{Azimov:2006he} has derived the even smaller value of the
$K^*N\Theta^+$ tensor coupling constant, employing the vector meson
dominance with SU(3) symmetry.  Note that the vector coupling constant
for the $K^*N\Theta^+$ vertex vanishes in SU(3) symmetry due to the
generalized Ademollo-Gatto theorem~\cite{Ledwig:1900ri}.  In
Ref.~\cite{Ledwig:1900ri}, the present authors have also shown that
the vector and tensor coupling constants for the $K^*N\Theta^+$ vertex
are very small within the framework of the chiral quark-soliton model
($\chi$QSM), taking into account SU(3) symmetry breaking effects.
Note that exactly the same self-consitent formalism and numerical
methods are used in the present work.  With this formalism used, we
can describe pentaquark observables on the same ground as those for
the octet baryons without doing any additional conceptual changes and
using the same parameter set. 

In addition to the $\Theta^+$, a recent GRAAL 
experiment~\cite{Kuznetsov:2004gy,Kuznetsov:2006de,Kuznetsov:2006kt} 
has announced a new finding of the nucleon-like resonance around
$1.67$ GeV in the neutron channel, measuring the cross section of
$\eta$ photoproduction off the deutron.  The corresponding width was
found to be around $40$ MeV, which probably results from a small
width being enlarged by Fermi-motion.  It was also shown in
Refs.~\cite{Kuznetsov:2004gy,Kuznetsov:2006de, Kuznetsov:2006kt} that
the resonant structure was not seen in the quasi-free proton channel.
This new resonance is consistent with the theoretical predictions by
Ref.~\cite{Diakonov:2003jj,Arndt:2003ga} of a new exotic nucleon like
state in that mass region.  In fact, the narrow width and its
dependence on the initial isospin state are the typical 
characteristics for the photo-excitation of the non-strange
anti-decuplet pentaquark~\cite{Polyakov:2003dx,Kim:2005gz}.  Very
recently, a new analysis of the free proton GRAAL
data~\cite{Kuznetsov:2007dy,Kuznetsov:2008gm} has been carried out,
the beam asymmetry being emphasized, and has revealed a resonance
structure with a mass around 1685 MeV with a width $\Gamma\leq 15$
MeV.  Note, however, that the results of Ref.~\cite{Kuznetsov:2007dy}
do not agree with those of Ref.~\cite{Bartalini:2007fg}.  For a
detailed discussion of this discrepancy, we refer to
Ref~\cite{Kuznetsov:2008gm}.  Furthermore, the LNS-GeV-$\gamma$
collaboration~\cite{tohoku,tohoku2} reports a new resonance at $1670$
MeV with a width $\Gamma\leq 50$ MeV in the $\gamma d\to \eta pn$
reaction.  It is consistent with the above-mentioned observation that
this resonance is enhanced in the $\gamma n\to \eta n$ reaction, while
it was not observed in the quasi-free proton channel as in the case of
Refs.\cite{Kuznetsov:2004gy,Kuznetsov:2006de,Kuznetsov:2006kt}.
Moreover, the CB-ELSA collaboration~\cite{CBELSA} has reported an
evidence for this nucleon-like resonance compatible with those of
GRAAL and LNS-GeV-$\gamma$, which are studied theoretically in
Ref~\cite{Fix:2007st}.  All these experimental facts are consistent
with the results for the transition magnetic moments in the
$\chi$QSM~\cite{Polyakov:2003dx,Kim:2005gz} as well as with the
phenomenological analysis for the non-strange pentaquark
baryons~\cite{Azimov:2005jj}.  Furthermore, recent theoretical 
calculations of the $\gamma N\to \eta N$
reaction~\cite{Choi:2005ki,Choi:2007gy} describe qualitatively 
well the GRAAL data, based on the values of the magnetic
transition moments in Refs.~\cite{Kim:2005gz,Azimov:2005jj}.
References~\cite{Arndt:2003ga,non_strange_partner2} have also
investigated the anti-decuplet focusing on the non-strange partners of
the $\Theta^+$, results of which are comparable with those in this
work. Due to all these experimental and phenomenological
results, we assume in the present work the anti-decuplet pentaquarks
to exist.  

Since the first prediction~\cite{Diakonov:1997mm} for the small width
of the $\Theta^+$, several calculations have been elaborated within
the same framework, i.e. in the $\chi$QSM in order to understand the
narrow decay width of the pentaquarks.  A formulation of the $\chi$QSM
in the light-cone framework various quark components in Fock space
have been decomposed~\cite{IMF_diakonov}: In the chiral limit a decay
width of around $2$ MeV was derived~\cite{IMF_imp_width}.  A
``model-independent approach'' as in Ref.~\cite{Diakonov:1997mm} was
extended to the axial-vector channel, based on the experimental data
of SU(3) baryon semileptonic decay constants and on the singlet
axial-vector constant, and has produced the decay width of the
$\Theta^+$ even smaller than 1 MeV~\cite{modelindep_theta_width}.
However, while the ``model-independent'' approach is plausible in
describing the smallness of the $\Theta^+$ decay width, one has to
understand the origin of its small width. 

In the present work, we want to investigate the axial-vector
transition constants and the widths of the baryon antidecuplet within
the framework of the $\chi$QSM with the symmetry-conserving
quantization~\cite{Praszalowicz:1998jm} employed, considering the 
rotational $1/N_c$ corrections and effects of SU(3) symmetry
breaking.  In contrast to the ``model independent'' approach of the 
$\chi$QSM in Ref.\cite{modelindep_theta_width}, the present
calculations are based on a self-consistent calculation of the
solitonic profile function without any refitting of $\chi$QSM
parameters.  The $\chi$QSM has been proved very successful not only in  
predicting the $\Theta^+$ but also even more in describing various
properties of SU(3) baryon octet and decuplet such as the mass
splittings, form factors and parton- and
antiparton-distributions~\cite{Christov:1995vm,Dressler:2001, 
GoekePPSU:2001,SchweitzerUPWPG:2001,OssmannPSUG:2005,KimPPYG:2005,
SilvaKUG:2005,WakamatsuN:2006,Wakamatsu:2005,Wakamatsuref:2003,
Wakamatsu:2000} and fulfilling all relevant sum rules of these
observables.  In particular, the dependence of almost all form 
factors on the momentum transfer is well reproduced within the
$\chi$QSM.  As a result, the strange electromagnetic form
factors~\cite{Goeke:2006gi} and the parity-violating asymmetries
of polarized electron-proton scattering, which require nine
different form factors (six electromagnetic form factors
$G_{E,M}^{(u,d,s)} (Q^2)$ and three axial-vector form factors
$G_{A}^{(u,d,s)} (Q^2)$), are in good agreement with experimental
data with one set of fixed parameters~\cite{Silva:2005qm}.  Thus, in
the present work, we extend the self-consistent $\chi$QSM to study the
axial-vector properties of the $\Theta^+$ and of the other members of
the baryon anti-decuplet. In this calculation, the resulting decay
width of the $\Theta^{+}$ will be shown to be $0.71\,\textrm{MeV}$. 

The structure of this work is as follows. In Section II, we present 
briefly the general formalism for the decay widths of the SU(3) baryons.  In
Section III, we review the $\chi$QSM and show how to calculate the
axial-vector transition constants.  We also explain how to derive the
widths of the baryon antidecuplet within the $\chi$QSM.  In Section
IV, we discuss the numerical results.  In the final Section, we
summarize the present work and draw conclusions.  Useful formulae are
given in the Appendix.

\section{Strong Decays of  the SU(3) baryons}
In the present Section, we briefly review the decay widths of the
SU(3) baryons, based on the effective Lagrangian approach.
Let us first consider the following decay modes:
\begin{equation}
  \label{eq:decay1}
\Delta\to N+\pi,\;\;\;  \Theta^{+}\to N+K,\;\;\;
N^{*}\to N+\pi(\eta),\;\;\;  N^{*}\to\Delta+\pi,
\end{equation}
where $\Delta$ denotes the $\Delta$ isobar, $N$ stands for the octet
nucleon, $\pi$, $K$, and $\eta$ are the pion, kaon, and $\eta$ meson,
respectively, and $N^{*}$ designates the anti-decuplet nucleon.  In
order to describe the above-given decays, we employ the following
effective Lagrangian:
\begin{eqnarray}
\mathcal{L}_{\Theta NK}&=& ig_{\Theta NK} \overline{\Theta}
\gamma^{5}KN + \mathrm{h.c.},\;\;\;\;\;\;\;
\mathcal{L}_{\Delta N\pi} = \frac{g_{\Delta N\pi}}{M_\Delta + M_N}
\overline{\Delta}_{\mu} (\partial^{\mu}\pi^{0}) N + \mathrm{h.c.} \cr 
\mathcal{L}_{N^{*}N\pi} &=& ig_{N^{*}N\pi}\overline{N^{*}}\gamma^{5}
\pi^{0} N + \mathrm{h.c.},\;\;\;
\mathcal{L}_{N^{*}\Delta\pi} = \frac{g_{N^{*}\Delta\pi}}{M_{N^*} +M_\Delta}
\overline{N^{*}}(\partial^{\mu} \pi^{0})\Delta_{\mu}+\mathrm{h.c.}, 
\end{eqnarray}
where $g_{BBM}$ are the strong coupling constants for the
baryon-baryon-meson vertices.  In the rest frame of a decaying
particle we have the formula for the strong decay as follows: 
\begin{eqnarray}
\Gamma(B_{1}\to B_{2}+M) & = & \frac{4\pi}{32\pi^{2}}
\frac{\mid{\bm k}_{M}\mid}{2s_{1}+1} \frac{1}{M_{1}^{2}}
\sum_{s_{1},s_{2}}|\mathcal{M}_{BBM}|^{2}\,\,\,,
\end{eqnarray}
where $M_{1}$ is the mass of the decaying particle, and $|{\bm k}_{m}|$
the three momentum of the meson. The $s_{1}$ and $s_{2}$ denote the third  
components of the spin for the initial and final baryons, 
respectively.  The invariant matrix elements $\mathcal{M}_{BBM}$ are 
therefore written as: 
\begin{eqnarray}
\mathcal{M}_{\Theta NK} &=& ig_{\Theta  NK}\, \overline{u}_{N}
(p_{N},s_{N}) \gamma^{5}u_{\Theta}(p_{\Theta},s_{\Theta}),\cr
\mathcal{M}_{\Delta N\pi} &=& \frac{g_{\Delta N\pi}}{M_\Delta + M_N}\, \overline{u}
(p_{N},s_{N})\, k_{\pi}^{\mu}\,u_{\mu}(p_{\Delta},s_{\Delta}),\cr 
\mathcal{M}_{N^{*}N\pi} &=& ig_{N^{*}N\pi}\, \overline{u}_{N}
(p_{N},s_{N}) \gamma^{5} u_{N^{*}}(p_{N^{*}},s_{N^{*}}),\cr
\mathcal{M}_{N^{*}\Delta\pi} &=& \frac{g_{N^{*}\Delta\pi}}{M_{N^{*}}+M_\Delta}
\overline{u}_{\mu}(p_{\Delta},s_{\Delta})\,k_{\pi}^{\mu}\,
u_{N^{*}}(p_{N^{*}},s_{N^{*}}). 
\end{eqnarray} 
Using the results in Appendix~\ref{sec:Decay-forumlae}, we obtain 
the decay widths for the processes $B_{1}\to B_{2}+M$ as follows:
\begin{eqnarray}
\Gamma(B_{\overline{10}}\to B_{8}+M) & = &
\frac{g_{B_{\overline{10}}B_{8}M}^{2}}{
M_{\overline{10}}^{2}}\,\,\frac{\mid{\bm k}_{M}\mid}{8\pi}\Big((
M_{\overline{10}}-M_{8})^{2}-m_M^{2}\Big),\\
\Gamma(B_{10}\to B_{8}+\pi^0) & = &
\frac{g_{B_{10}B_{8}\pi}^{2}}{(M_{10} + M_8)^2 M_{10}^{2}}
\,\,\frac{|{\bm
    k}_{\pi}|^3}{24\pi}\,\,\Big((M_{8}+M_{10})^{2}-m_{\pi}^{2}\Big),\\ 
\Gamma(N^{*}\to\Delta+\pi^0) & = &
\frac{g_{N^{*}\Delta\pi}^{2}}{(M_{N^*}+M_\Delta)^2 M_{\Delta}^{2}}\,\,\frac{|{\bm
    k}_{\pi}|^3}{12\pi}
\,\,\Big((M_{N^{*}}+M_{\Delta})^{2}-m_{\pi}^{2}\Big),\,\,\, 
\label{decay_formula}
\end{eqnarray}
with the meson momentum 
\begin{equation}
  \label{eq:mesonmom1}
{\bm k}_{M}^{2} = \frac{(M_{1}^{2}-M_{2}^{2}
  + m_M^{2})^{2}-4M_{1}^{2}m_M^{2}}{4M_{1}^{2}}.  
\end{equation}

Using the generalized Goldberger-Treiman
relations (see Eq.(\ref{GT_generalized}) given below), we can relate 
the axial-vector transition constants to the strong coupling
constants.  In the case of the $B_{\overline{10}}\to B_{8}$
transitions, the axial-vector form factors are generally defined as
the following matrix elements for the axial-vector current:
\begin{eqnarray}
  &&\langle B_{8}^{\prime}(p^{\prime},\,s')|A^{\mu
   a}(0)|B_{\overline{10}}(p,\,s)\rangle = \langle
 B_{8}^{\prime}(p^{\prime}\,s')|\overline{\psi}(0)\gamma^{\mu}\gamma^{5}
 \frac{\lambda^{a}}{2}\psi(0)|B_{\overline{10}}(p,\,s)\rangle\cr
 & = &
 \overline{u}_{8}(p^{\prime},\,s')[G_{A}(Q^{2})\gamma^{\alpha} +
 G_{P}(Q^{2}) q^{\alpha} + G_{T}(Q^{2}) P^{\alpha}] \gamma^{5} 
\frac{\lambda^{a}}{2}u_{\overline{10}}(p,\,s),
\label{axialspin1}
 \end{eqnarray}
where $Q^{2}=-q^{2}=-(p^{\prime}-p)^{2}$ and $P=p^{\prime}+p$.  The
$\lambda^{a}$ denotes the SU(3) flavor Gell-Mann matrices, satisfying
$\{\lambda^a,\,\lambda^b\}=2\delta^{ab}$.  The axial-vector
transition constant is defined as the value of 
$G_{A}(Q^{2})$ at $Q^{2}=0$, i.e. $g_{A}^{*}=G_{A}(0)$. The
generalized Goldberger-Treiman relation connects the axial-vector
transition constant to the corresponding strong coupling constant as
follows:  
\begin{equation}
g_{B_{1}B_{2}M}=\frac{g_{A}^{*}\,\,(M_{1}+M_{2})}{2f_{M}},
\label{GT_generalized}
\end{equation}
where $f_M$ stands for the corresponding meson decay constant.  

As for the transitions from the baryon decuplet to the octet, we need
to deal with the Lorentz structure of spin-$3/2$ particles, so that
we have more form factors, i.e. Adler form
factors~\cite{Delta_Hemmert,Delta_Zhu,Delta_lattice,Adler_FF}.  The 
axial-vector transition for the $\Delta^{+}\to p+\pi^{0}$ process is
then expressed in terms of four independent form factors:
\begin{eqnarray}
\langle\Delta^{+}(p',\,s')|A^{\mu 3}(0)|p(p,\,s)\rangle
& = & \langle \Delta^{+}(p',\,s')| \overline{\psi}(0)
\gamma^{\mu} \gamma^{5}\frac{\lambda^{3}}{2}\psi(0)|p(p,\,s)\rangle \cr
& = & 
\overline{u}_{\nu}^{\Delta^+}(p^{\prime},s^{\prime})\Big[C_{5}^{A}(Q^{2})
g^{\mu\nu}+C_{6}^{A}(Q^{2})q^{\mu}q^{\nu}\cr 
& + & 
\Big\{C_{3}^{A}(Q^{2})\gamma_{\lambda} + C_{4}^{A}(Q^{2})
p_{\lambda}^{\prime}\Big\}( q^{\lambda}g^{\mu\nu} -
q^{\nu}g^{\lambda\mu})\Big]u_p (p,s),
\label{Adler}
\end{eqnarray}
where $u_{\nu}^{\Delta^+}$ and $u_p$ denote the Rarita-Schwinger and
Dirac spinors for the $\Delta$ and proton, respectively.  In this
case, the axial-vector transition constant is defined as
$C_{A}^{*}=C_{5}^{A}(0)$.  In order to derive the Goldberger-Treiman
relation for the spin-3/2 baryon, we first determine the divergence of
the axial-vector current~\cite{CHP}, which should vanish in the chiral 
limit:  
\begin{equation}
i\overline{u}_{\nu}^{\Delta^+}(p^{\prime},s^{\prime}) q^{\nu}\Big[
C_{5}^{A}(q^{2}) + C_{6}^{A}(q^{2})q^{2} \Big]u_p (p,s) =
0.  
\label{eq:da}
\end{equation}
Thus, we obtain the following relation:
\begin{equation}
  \label{eq:da2}
C_{5}^{A}(q^{2}) + C_{6}^{A}(q^{2})q^{2} = 0.   
\end{equation}
Since we have $C_{5}^{A}(0)\neq0$, the term $C_{6}^{A}(q^{2})$
must have a pole at $q^{2}=0$.  We can identify the pole term in
Eq.(\ref{eq:da}) in the following way:
\begin{eqnarray}
\overline{u}_{\nu}^{\Delta^+}(p^{\prime},s^{\prime})
q^{\nu}\,C_{6}^{A}(q^{2})\,\, u_p (p,s)\,\,\, 
& \rightarrow & \,\,\, \frac{ g_{\Delta  N \pi}}{M_\Delta + M_N}  \,
\overline{u}_{\nu}^{\Delta^+}( p^{\prime}, s^{\prime}) \,
g^{\mu\nu}\,u_p (p,s)\,\frac{i}{q^{2}}\, if_{\pi}\, q_{\mu}, \\  
\lim_{q^{2}\rightarrow0}\Big(C_{5}^{A}(q^{2})+C_{6}^{A}(q^{2})q^{2}\Big)
& = & \lim_{q^{2}\rightarrow0}\left[C_{5}^{A}(q^{2})-\frac{g_{\Delta
      N\pi}}{(M_\Delta + M_N)
  q^{2}}f_{\pi}q^{2}\right] = 0.
\end{eqnarray} 
Therefore, we get the Goldberger-Treiman relation for spin-$3/2$ baryon 
\cite{Delta_Hemmert,Delta_Zhu,Delta_Goldberger} as follows:
\begin{equation} 
C_{5}^{A}(0)=f_{\pi}\,\, \frac{g_{\Delta N\pi}}{M_\Delta + M_N}.
\end{equation}
\section{The chiral quark-soliton model}
\subsection{General Formalism}
The SU(3) $\chi$QSM is characterized by the following partition
function in Euclidean space:
\begin{equation}
\label{eq:part}
\mathcal{Z}_{\mathrm{\chi QSM}} = \int \mathcal{D}\psi\mathcal{D}\psi^\dagger
 \mathcal{D}\pi\exp\left[-\int d^4 x\psi^\dagger D(\pi)\psi\right]
= \int \mathcal{D}\pi\exp(-S_{\mathrm{eff}}[\pi]),
\end{equation}
where $\psi$ and $\pi$ denote the quark and pseudo-Goldstone boson
fields, respectively.  The $S_{\mathrm{eff}}$ stands for the effective
chiral action expressed as 
\begin{equation}
  \label{eq:echl}
S_{\mathrm{eff}} = -N_c\mathrm{Tr}\ln D(\pi),
\end{equation}
where $\mathrm{Tr}$ represents the functional trace, $N_c$ the number
of colors, and $D$ the Dirac differential operator in Euclidean space:
\begin{equation}
 \label{eq:Dirac}
D(U) = \gamma_4(i\rlap{/}{\partial} -\hat{m} -MU^{\gamma_5}) = \partial_4
+ h(U) + \delta m.
\end{equation}
Here, the $\hat{m}$ denotes the current quark matrix $\hat{m} =
\mathrm{diag}(\overline{m},\,\overline{m},\,m_{\mathrm{s}})$, isospin
symmetry being assumed.  The $\partial_4$ designates the derivative
with respect to the Euclidean time and $h(U)$ stands for the Dirac
single-quark Hamiltonian:
\begin{equation}
h(U)=-i\gamma_4\gamma_i\partial_i + \gamma_4 MU^{\gamma_5}
+\gamma_4\overline{m}.
\label{eq:diracham}
\end{equation}
The $\delta m$ is the the matrix of the decomposed current quark
masses:
\begin{equation}
  \label{eq:deltam}
\delta m = M_1\gamma_4\bm 1 + M_8 \gamma_4\lambda^8,
\end{equation}
where $M_1$ and $M_8$ are singlet and octet components of the current
quark masses defined as $M_1 =(-\overline{m}+m_{\mathrm{s}})/3$ and
$M_8=(\overline{m}-m_{\mathrm{s}})/\sqrt{3}$.  The $\overline{m}$ is
the average of up- and down-quark masses.  The chiral field
$U^{\gamma_5}$ is written as
\begin{equation}
U^{\gamma_5} = \exp(i\gamma_5 \lambda^a\pi^a) =
\frac{1+\gamma_5}{2}U + \frac{1-\gamma_5}{2} U^\dagger
\end{equation}
with $U=\exp(i\lambda^a\pi^a)$.  We assume here Witten's embedding of SU(2) into SU(3):
\begin{equation}
  \label{eq:embed}
U_{\mathrm{SU(3)}} = \left(\begin{array}{lr} U_{\mathrm{SU(2)}} & 0
    \\ 0 & 1   \end{array} \right)
\end{equation}
with the SU(2) hedgehog chiral field
\begin{equation}
  \label{eq:hedgehog}
U_{\mathrm{SU2}}=\exp[i\gamma_5 \hat{\bm
  n}\cdot\bm\tau P(r)].
\end{equation}
In order to solve the partition function in
Eq.(\ref{eq:part}), we have to take the large $N_c$ limit and solve it
in the saddle-point approximation, which corresponds at the classical
level to finding the profile function $P(r)$ in
Eq.(\ref{eq:hedgehog}).  In fact, the profile function can be obtained
by solving numerically the functional equation coming from $\delta
S_{\mathrm{eff}}/\delta P(r) =0$, which yields a classical soliton
field $U_c$ constructed from a set of single quark energies $E_n$ and
corresponding states $|n\rangle$ related to the eigenvalue equation
$h(U)|n\rangle =E_n|n\rangle$.

Since the classical soliton does not have the quantum number of the
baryon states, we need to restore them by the semiclassical
quantization of the rotational and translational zero modes.  Note
that the zero modes can be treated exactly within the functional
integral formalism by introducing collective coordinates. The detailed
formalism can be found in Refs.~\cite{Christov:1995vm,Kim_eleff}.
Considering the rigid rotations and translations of the classical
soliton $U_c$, we can express the soliton field as
\begin{equation}
U(\bm x, t) = A(t)U_c(\bm x - \bm z(t))A^\dagger (t),
\end{equation}
where $A(t)$ denotes a unitary time-dependent SU(3) collective
orientation matrix and $\bm z(t)$ stands for the time-dependent
displacement of the center of mass of the soliton in coordinate
space.

Having introduced the zero modes as mentioned above, the
Dirac operator in Eq.(\ref{eq:Dirac}) is changed to the following
form:
\begin{equation}
D(U)=T_{z(t)}A(t) \left[ D(U_{c}) + i\Omega(t) -
  \dot{T}_{z(t)}^{\dagger}T_{z(t)}-i\gamma^{4}A^{\dagger}(t)\delta m
  A(t) \right]T_{z(t)}^{\dagger}A^{\dagger}(t),
\end{equation}
where the $T_{z(t)}$ denotes the translational unitary operator and
the $\Omega(t)$ represents the angular velocity of the soliton that is
defined as
\begin{equation}
\Omega=-iA^{\dagger}\dot{A}=-\frac{i}{2}\textrm{Tr} (
A^{\dagger}\dot{A}\lambda^{\alpha})\lambda^{\alpha}=\frac{1}{2}
\Omega_{\alpha}\lambda^{\alpha}.
\end{equation}
Assuming that the soliton rotates and moves slowly, we can treat the
$\Omega(t)$ and $\dot{T}_{z(t)}^{\dagger}T_{z(t)}$ perturbatively.
Moreover, since the flavor SU(3) symmetry is broken weakly, we can
also deal with $\delta m$ perturbatively.

Having quantized collectively, we obtain the following collective
Hamiltonian
\begin{equation}
  \label{eq:Ham}
H_{\textrm{coll}}=H_{\mathrm{sym}} + H_{\mathrm{sb}},
\end{equation}
where $H_{\mathrm{sym}}$ and $H_{\mathrm{sb}}$ represent the SU(3)
symmetric and symmetry-breaking parts, respectively:
\begin{eqnarray}
H_{\mathrm{sym}} &=& M_{c} + \frac{1}{2I_{1}}\sum_{i=1}^3 J_{i}J_{i} +
\frac{1}{2I_{2}} \sum_{a=4}^7 J_{a} J_{a} + \frac{1}{\overline{m}} M_{1}
\Sigma_{SU(2)} ,\cr
H_{\mathrm{sb}} &=& \alpha D_{88}^{(8)}(A) + \beta Y +
\frac{\gamma}{\sqrt{3}} D_{8i}^{(8)}(A)J_{i}.
\end{eqnarray}
The $M_c$ denotes the mass of the classical soliton and $I_{i}$ and
$K_{i}$ are the momenta of inertia of the
soliton~\cite{Christov:1995vm}, of which the corresponding
expressions can be found in Ref.~\cite{Blotz_mass_splittings}
explicitly.  The components $J_i$  denote the spin generators and
$J_a$ correspond to those of right rotations in flavor SU(3) space.  The
$\Sigma_{\mathrm{SU(2)}}$ is the SU(2) pion-nucleon sigma term. The
$D_{88}^{(8)}(A)$ and $D_{8i}^{(8)}(A)$ stand for the SU(3) Wigner $D$
functions in the octet representation.  The $Y$ is the hypercharge
operator.  The parameters $\alpha$, $\beta$, and $\gamma$ in the
symmetry-breaking Hamiltonian are expressed, respectively, as follows: 
\begin{equation}
\alpha = \frac{1}{\overline{m}} \frac{1}{\sqrt{3}} M_{8}
\Sigma_{SU(2)} - \frac{N_{c}}{\sqrt{3}} M_{8} \frac{K_{2}}{I_{2}},
\;\;\;\;
\beta = M_{8} \frac{K_{2}}{I_{2}}\sqrt{3},\;\;\;\;
\gamma = -2\sqrt{3}M_{8}\left(\frac{K_{1}}{I_{1}} -
  \frac{K_{2}}{I_{2}}\right).
\end{equation}
The collective wave-functions of the Hamiltonian in Eq.(\ref{eq:Ham})
can be found as SU(3) Wigner $D$ functions in representation
$\mathcal{R}$:
\begin{equation}
  \label{eq:Wigner}
\langle A|\mathcal{R},B(YII_{3},Y^{\prime}JJ_{3}) \rangle =
\Psi_{(\mathcal{R}^{*};Y^{\prime}JJ_{3})}^{(\mathcal{R};YII_{3})}(A) =
\sqrt{\textrm{dim}(\mathcal{R})}\,(-)^{J_{3}+Y^{\prime}/2}\,
D_{(Y,I,I_{3})(-Y^{\prime},J,-J_{3})}^{(\mathcal{R})*}(A).
\end{equation}
The $Y'$ is related to the eighth component of the angular velocity
$\Omega$ that is due to the presence of the discrete valence quark
level in the Dirac-sea spectrum.  Its presence has no effect on the
chiral field, so that it is constrained to be $Y'=-N_c/3=-1$.  In fact,
this constraint allows us to have only the SU(3) representations with
zero triality.

The effects of flavor SU(3) symmetry breaking having been taken into
account, the collective baryon states are not in a pure
representation but start to get mixed with other representations.
This can be treated by considering the first-order perturbation for 
the collective Hamiltonian:
\begin{equation}
|B_{\mathcal{R}}\rangle = |B_{\mathcal{R}}^{\mathrm{sym}} \rangle -
\sum_{\mathcal{R}^{\prime}\neq\mathcal{R}} |B_{\mathcal{R}^{\prime}}
\rangle \frac{\langle B_{\mathcal{R}^{\prime}}|\,
  H_{\textrm{sb}}   \,|B_{\mathcal{R}}\rangle}{M(\mathcal{R}^{\prime}) -
  M(\mathcal{R})}\,.
\label{wfc}
\end{equation}
Solving Eq.(\ref{wfc}), we obtain the collective wave functions for
the baryon octet, decuplet, and anti-decuplet:
\begin{eqnarray}
|B_{8}\rangle & = & |8_{1/2},B\rangle + c_{\overline{10}}^{B}
|\overline{10}_{1/2}, B\rangle + c_{27}^{B}|27_{1/2},B\rangle,
\label{B8}\\
|B_{10}\rangle & = & |10_{3/2},B\rangle+a_{27}^{B}|27_{3/2},B\rangle +
a_{35}^{B}|35_{3/2},B\rangle,
\label{B10}\\
|B_{\overline{10}}\rangle & = & |\overline{10}_{1/2},B\rangle +
d_{8}^{B} |8_{1/2},B\rangle + d_{27}^{B}|27_{1/2},B\rangle +
d_{\overline{35}}^{B}| \overline{35}_{1/2},B\rangle,
\label{wavefunctions}
\end{eqnarray}
with the mixing coefficients
\begin{eqnarray}
c_{\overline{10}}^{B}&=&c_{10}\left(\begin{array}{c}
\sqrt{5}\\
0\\
\sqrt{5}\\
0\end{array}\right),\;\; c_{27}^{B}=c_{27}\left(\begin{array}{c}
\sqrt{6}\\
3\\
2\\
\sqrt{6}\end{array}\right),
\nonumber \\
a_{27}^{B}& = & a_{27}\left(\begin{array}{c}
\sqrt{15/2}\\
2\\
\sqrt{3/2}\\
0\end{array}\right),\;\; a_{35}^{B}=a_{35}\left(\begin{array}{c}
5/\sqrt{14}\\
2\,\sqrt{5/7}\\
3\,\sqrt{5/14}\\
2\,\sqrt{5/7}\end{array}\right),
\nonumber \\
d_{8}^{B} & = & d_{8}\,\left(\begin{array}{c}
0\\
\sqrt{5}\\
\sqrt{5}\\
0\end{array}\right),\;\;  d_{27}^{B} \;=\;
d_{27}\,\left(\begin{array}{c} 0\\
\sqrt{3/10}\\
2/\sqrt{5}\\
\sqrt{3/2}\end{array}\right),\;\; d_{\overline{35}}^{B}=\,
d_{\overline{35}}\,\left(\begin{array}{c} 
1/\sqrt{7}\\
3/(2\sqrt{14})\\
1/\sqrt{7}\\
\sqrt{5/56}\end{array}\right),
\end{eqnarray}
in the bases $[N,\Lambda,\Sigma,\Xi]$,
$[\Delta,\Sigma_{10}^{*},\Xi_{10}^{*},\Omega]$, and
$[\Theta,N_{\overline{10}}^{*},\Sigma_{\overline{10}}^{*},
\Xi_{\overline{10}}^{*}]$, respectively.  The coefficients $c_i$,
$a_i$ and $d_i$ are expressed as 
\begin{eqnarray}
c_{\overline{10}} & = & -\frac{I_{2}}{15}\Big(\alpha +
\frac{1}{2}\gamma\Big),\;\; 
c_{27}=-\frac{I_{2}}{25}\Big(\alpha-\frac{1}{6}\gamma\Big),
\nonumber \\
a_{27} & = & -\frac{I_{2}}{8}\Big(\alpha + \frac{5}{6} \gamma
\Big),\;\; a_{35}=-\frac{I_{2}}{24}\Big(\alpha - \frac{1}{2} \gamma
\Big),  
\nonumber \\
d_{8} & = & \;\;\frac{I_{2}}{15}\Big(\alpha+\frac{1}{2}\gamma\Big),\;\;
d_{27} = -\frac{I_{2}}{8}\Big(\alpha - \frac{7}{6} \gamma \Big),\;\;  
d_{\overline{35}} = -\frac{I_{2}}{4} \Big(\alpha + \frac{1}{6} \gamma
\Big). 
\label{mixing_coeff}
\end{eqnarray}

Now, we are in a position to evaluate the baryonic matrix elements
given in Eqs.(\ref{axialspin1},\ref{Adler}) within the framework of the
$\chi$QSM.  In general, the baryonic matrix element of the
axial-vector current
$A_{\mu}^{a} = i {\psi}^\dagger \gamma_\mu \lambda^a/2 \psi$
can be expressed as the following correlation function in the
functional integral:
\begin{eqnarray}
\langle B^{\prime}(p^{\prime})|A_{\mu}^{a} |B(p)\rangle & = &  
\frac{1}{\mathcal{Z}} \lim_{T\to\infty} e^{-ip_{4}^{\prime}
  \frac{T}{2}+ip_{4}\frac{T}{2}} \int d^{3}x^{\prime}\,d^{3}x\;
e^{i{\bm p}\cdot{\bm x} - i{\bm p}^{\prime}\cdot{\bm x}^{\prime}}
\cr
 &  & \hspace{-3cm} \times\int\mathcal{D} \psi^{\dagger}\mathcal{D}\psi
\mathcal{D} U
J_{B^{\prime}}\left(\frac{T}{2},{\bm x}^{\prime}\right)\,A_{\mu}^{a}
(0) J_{B}^{\dagger}\left(-\frac{T}{2},{\bm x}\right)\,
 e^{-\int d^{4}x\,\psi^\dagger D(U)\psi}.
\label{eq:corr}
\end{eqnarray}
with the baryonic current that consists of $N_c$ quarks and the baryon
state:
\begin{eqnarray}
J_B(x) &=& \frac{1}{N_c!}\epsilon_{i_1\cdots
  i_{N_c}}\Gamma_{JJ_3TT_3Y}^{\alpha_1\cdots\alpha_2}\psi_{\alpha_1i_1}
(x)\cdots \psi_{\alpha_{N_c}i_{N_c}}(x),\cr
|B(p)\rangle & = &
\lim_{x_{4}\to-\infty}\,\frac{1}{\sqrt{\mathcal{Z}}}\,
e^{ip_{4}x_{4}}\,\int d^{3} x \, e^{i\,{\bm p}\cdot{\bm x}}\,
J_{B}^{\dagger}(x)\,|0\rangle.
\end{eqnarray}
Here, $\alpha_1\cdots\alpha_{N_c}$ denote spin-flavor indices,
whereas $i_1\cdots i_{N_c}$ represent color indices.

We can solve Eq.(\ref{eq:corr}) in the saddle-point approximation
justified in the large $N_c$ limit.  In this approximation and with
the help of the zero-mode quantization, the functional integral over
the chiral field turns out to be the integral over the rotational zero
modes.  Since we will consider the rotational $1/N_c$ corrections and
linear $m_{\mathrm{s}}$ corrections, we expand the quark propagators
in Eq.(\ref{eq:corr}) with respect to $\Omega$ and $\delta m$ to the
linear order and $\dot{T}_{z(t)}^{\dagger}T_{z(t)}$ to the zeroth
order.

Having carried out a tedious but straightforward
calculation~(see Refs.~\cite{Christov:1995vm,Kim_eleff} for details),
we finally can express the baryonic matrix element such as
Eq.(\ref{axialspin1}) as a Fourier transform in terms of the
corresponding quark densities and collective wave-functions of the
baryons:
\begin{equation}
\langle B^{'}(p^{'})| A_{\mu}^{a} (0) |B(p)\rangle =
\int d A \int d^{3}z\,\, e^{i{\bm q}\cdot{\bm z}}\,
\Psi_{B^{\prime}}^{*}(A) \mathcal{F}_{\mu}^a({\bm z})\Psi_{B}(A),
\label{eq:model}
\end{equation}
where $\Psi(A)$ denote the collective wave-functions and
$\mathcal{F}_{\mu}^{a}$ represents the quark densities corresponding
to the current operator $A_\mu^a$. 

Following the formalism presented above, we arrive at the final
expressions for the axial-vector form factors:
\begin{eqnarray}
G_{A}(Q^{2}) & = & G_{A}^{(\Omega^{0},m_{s}^{0})}(Q^{2}) +
G_{A}^{(\Omega^{1},m_{s}^{0})}(Q^{2}) +
G_{A}^{(m_{s}^{1}),\textrm{op}}(Q^{2})+G_{A}^{(m_{s}^{1}),\textrm{wf}}(Q^{2}),
\label{eq:cqsmff}
\end{eqnarray} 
where the first term corresponds to the leading order
($\Omega^{0},m_{s}^{0}$), the second one to the first $1/N_{c}$
rotational correction ($\Omega^{1},m_{s}^{0}$) and the later to linear
$m_{s}$ corrections coming from the operator and the wave-function
corrections, respectively. 

In the $\chi$QSM Hamiltonian of Eq.(\ref{eq:diracham}) the constituent
quark mass $M$ is the only free parameter and $M=420\,\textrm{MeV}$
is known to reproduce very well experimental
data~\cite{SilvaKUG:2005,ChristovEMFF,Silva:2005qm,Kim_eleff}.  
Though the $M=420$ MeV yields the best results for the baryon octet, 
we will present also those for $M=400\,\textrm{MeV}$ and
$M=450\,\textrm{MeV}$ to see the $M$ dependence of the results in this
work.  Throughout this work the strange current quark mass is fixed to
$m_{\mathrm{s}}=180\,\textrm{MeV}$.  In order to tame the divergent
quark loops, we employ in this work the proper-time regularization.
The cut-off paramter and $\overline{m}$ are fixed for a given $M$
to the pion decay constant $f_{\pi}$ and $m_\pi$.  The numerical
results for the moments of inertia and mixing coefficients are
summarized in Table~\ref{Numerical-values} for $M=420$ MeV. 
\begin{table}[t]
\caption{\label{Numerical-values} Moments of inertia and mixing
coefficients for $M=420\,\textrm{MeV}$.}
\begin{center}
\begin{tabular}{ccccccccccc}
\hline
$I_{1}\,[\textrm{fm}]$&
$I_{2}\,[\textrm{fm}]$&
$K_{1}\,[\textrm{fm}]$&
$K_{2}\,[\textrm{fm}]$&
$\Sigma_{\pi N}\,[\textrm{MeV}]$ &
$c_{\overline{10}}$&
$c_{27}$&
$d_{8}$&
$d_{27}$&
$a_{27}$&
$a_{35}$
\\ \hline
$1.06$&
$0.48$&
$0.42$&
$0.26$&
$41$&
$0.037$&
$0.019$&
$-0.037$&
$0.043$&
$0.074$&
$0.018$
\tabularnewline
\hline
\end{tabular}
\end{center}
\end{table}
The results in Table~\ref{Numerical-values} are obtained with
the same paramters used in previous works
\cite{Silva:2005qm,silva_data:1999,silva_data:2002}.    
We want to emphasize that all model parameters are the same as
before.  In previous works, the axial-vector form factors for the
nucleon were already calculated.  The axial-vector constants
$g_{A}^{3},g_{A}^{8}$ were found to be $g_{A}^{3}=1.176$ and
$g_{A}^{8}=0.36$ which is in very good agreement with experimental
data $g_{A}^{3}=1.267\pm0.0029$ \cite{PDG:2006} and
$g_{A}^{8}=0.338\pm0.15$ \cite{exp_axial_data_g8}. 

\subsection{Axial-Vector Form Factors in the $\chi$QSM}
The axial-vector form factors for baryons are generally expressed
in terms of the quark matrix elements given in Eq.(\ref{axialspin1}).
Since we are using an explicit self-consistent soliton profile
derived from an action principle minimizing the nucleon
energy, we calculate $G_A(Q^2)$ via the baryonic matrix element.  In
order to extract the form factor $G_{A}(Q^{2})$ it is helpful to make
the vector products to the spacial component of this current  
by the vector ${\bm q}$ and to perform an average over the angular
momentum transfer orientation, $\int\frac{d\Omega_{q}}{4\pi}$.  Taking
the rest frame for the initial baryon ($p=(M_{B},0)$,
$p^{\prime}=(E_{B^{\prime}},-{\bm q})$), we get  
\begin{eqnarray}
\int\frac{d\Omega_{q}}{4\pi} {\bm q}\times\Big({\bm q}\times\langle
B_{s^{\prime}}^{\prime}(p^{\prime})|{\bm A} |B_{s}(p)\rangle\Big) &
= & -\frac{2}{3}{\bm q}^{2}
\sqrt{\frac{E_{B^{\prime}}+M_{B^{\prime}}}{2M_{B^{\prime}}}} 
 G_{A}(Q^{2})\,\phi_{s^{\prime}}^{\dagger}
 {\bm\sigma}\phi_{s}.
\label{eq:pa}\end{eqnarray}
Choosing equal initial and final baryon spins and using
Eq.(\ref{eq:corr}) with Eq.(\ref{eq:pa}), we derive from the third 
spacial component the $\chi$QSM expression for the axial-vector
constant as follows~\cite{Silva:2005qm}:
\begin{eqnarray}
G_{A}^{a}(0) & = & \int d^{3}z\, \langle
B^{\prime}_{\overline{10}}\uparrow|\mathcal{G}_{A}^{a}({\bm
  z})|B_8\uparrow\rangle. 
\label{eq:axial-model}
\end{eqnarray}
The axial-vector density $\mathcal{G}_{A}^{a}({\bm z})$ for a
certain flavor part $a$ is given by
\begin{eqnarray}
\mathcal{G}_{A}^{a}({\bm z}) & = & -\sqrt{\frac{1}{3}}
D_{a3}^{(8)} \mathcal{A}({\bm z}) + \frac{1}{3\sqrt{3}}
\frac{1}{I_{1}} D_{a8}^{(8)} J_{3} \mathcal{B}({\bm z}) -
\sqrt{\frac{1}{3}} \frac{1}{I_{2}}D_{ap}^{(8)}J_{q}d^{pq3}
\mathcal{C}({\bm z}) 
\nonumber \\ 
 & - & \frac{1}{3\sqrt{2}} \frac{1}{I_{1}} D_{a3}^{(8)}
 \mathcal{D}({\bm z}) - \frac{2}{3\sqrt{3}} \frac{K_{1}}{I_{1}} M_{8}
 D_{83}^{(8)} D_{a8}^{(8)} \mathcal{B}({\bm z})
\nonumber \\
 & + &  \frac{2}{\sqrt{3}}
 \frac{K_{2}}{I_{2}}\, M_{8}\, D_{8p}^{(8)}D_{a q}^{(8)}d^{pq3}
 \mathcal{C}({\bm z}) - \frac{2}{\sqrt{3}} \Big[M_{1}D_{a3}^{(8)} +
 \frac{1}{\sqrt{3}} 
 M_{8}D_{88}^{(8)} D_{a3}^{(8)} \Big] \mathcal{H}({\bm z})
\nonumber \\
 & + &  \frac{2}{3\sqrt{3}}M_{8} D_{83}^{(8)} D_{a8}^{(8)}
 \mathcal{I}({\bm z}) - \frac{2}{\sqrt{3}}M_{8} D_{8p}^{(8)} D_{a
   q}^{(8)}d^{pq3} \mathcal{J}({\bm z})\,\,\,,
\label{eq: axial-densities thesis}
\end{eqnarray}
where the densities $\mathcal{A}({\bm z}),\mathcal{B}({\bm z}),\cdots$ 
are given in Appendix~\ref{app:axdens}.  In the case of the Adler form
factors Eq.(\ref{Adler}) we have 
\begin{eqnarray}
 &  & \int\frac{d\Omega_{q}}{4\pi} \Big[{\bm q}\times({\bm q} \times 
 \langle \Delta^{+}(p^{\prime},\,s')|{\bm A}^{a=3}|p(p,\,
 s)\rangle\Big]_{z}
\nonumber \\
 & = &
 \int \frac{d\Omega_{q}}{4\pi} \Big[{\bm q}\times\Big({\bm q} \times  
 \overline{u}_{\nu}^{\Delta^+} (p^{\prime},s') \Big[C_{5}^{A}(Q^{2})
 g^{k\nu} + C_{4}^{A}(Q^{2}) p_{\lambda}^{\prime}q^{\lambda}g^{k\nu}\Big]
 u_p(p,\,s) \hat{e}^{k} \Big)\Big]_{z},
\end{eqnarray}
where the form factor $C_{3}^{A}$ is taken to be
zero~\cite{Delta_Hemmert,Delta_lattice} and we can neglect 
$C_{4}^{A}$.  This treatment is similar to that for the 
$N^{*}(1440)\to\Delta\pi$ decay in Ref.\cite{N(1440)_decay}. 
The Rarita-Schwinger spinor $u^{k}(p^{\prime},\frac{1}{2})$ with
its third component $+1/2$ is expressed as follows~\cite{Pilkuhn}: 
\begin{equation}
u^{k}(p^{\prime},\frac{1}{2}) =
\frac{1}{\sqrt{3}}u(p^{\prime},-\frac{1}{2})
\varepsilon^{k}(+1)+\sqrt{\frac{2}{3}}u(p^{\prime},+\frac{1}{2}) 
\varepsilon^{k}(0) 
\end{equation}
\begin{equation}
\overline{u}^{k}(p^{\prime},\frac{1}{2})u(p,\frac{1}{2}) =
\sqrt{\frac{2}{3}} \sqrt{\frac{E_{N}+M_{N}}{2M_{N}}} \hat{e}^{k}, 
\end{equation} 
and we arrive at the expression for $C_5^a$:
\begin{eqnarray}
C_{5}^{a}(0) & = & \sqrt{\frac{3}{2}}\int d^{3}z\, \langle
B_{10}^{\prime} \uparrow|\mathcal{G}_{10}^{a}({\bm
  z})|B_{8} \uparrow\rangle,
\label{eq:delta-axial}
\end{eqnarray}
where $\mathcal{G}_{10}^{a}$ is the axial-vector density
for the $B_{\overline{10}}^{\prime}\to B_8$ transition.  

The baryonic matrix element of the $D$ functions 
can be expressed in terms of SU(3) Clebsch-Gordan
coefficients~\cite{CHP,deSwart}: 
\begin{eqnarray}
\langle B_{\mathcal{R}^{\prime}}^{\prime}|D_{a
  m}^{n}(A)|B_{\mathcal{R}}\rangle & = & \sqrt{ \frac{\textrm{dim}
    \mathcal{R}^{\prime}}{ \textrm{dim}\mathcal{R}}}(-1)^{\frac{1}{2}
  Y_{s}^{\prime}+S_{3}^{\prime}}(-1)^{\frac{1}{2}Y_{s}+S_{3}}\cr
&  & \times\sum_{\gamma}\left(\begin{array}{ccc}
\mathcal{R}^{\prime} & n & \mathcal{R}_{\gamma}\\
Q^{\prime} & a & Q\end{array}\right)\left(\begin{array}{ccc}
\mathcal{R}^{\prime} & n & \mathcal{R}_{\gamma}\\
-Y_{s}^{\prime}S^{\prime}-S_{3}^{\prime} & m & -Y_{S}S-S_{3}
\end{array}\right),
\end{eqnarray}
with $Q=YII_{3}$. 
The relevant results are listed in Appendix~\ref{app:bame}. \\

\section{Results and Discussion}
\subsection{Mass Splittings\label{chapter:mass-splittings}}
The $\chi$QSM in the present from is not able to calculate
absolute masses since rotational and translational quantum corrections
are not calculated~\cite{pusching_soliton}.  However, the mass
splittings are accessible.  The mass splittings between the baryon
octet and the antidecuplet have been already studied in
detail~\cite{Diakonov:1997mm,mixings:2004}.  Note that there have been
also some discussions about the applicability of the collective
quantization due to the rigid rotation of the chiral 
soliton~\cite{Cohen:2003yi,Cohen:2003mc,Pobylitsa:2003ju,
Diakonov:2003ei}.  The symmetry-breaking part $H_{\textrm{sb}}$ of the
Hamiltonian in Eq.(\ref{eq:Ham}) enables us to calculate baryon mass
splittings for various representions as done in 
\cite{Christov:1995vm,Blotz_mass_splittings}.  The mass splittings
between the baryon octet and the decuplet and anti-decuplet are 
given in terms of the soliton moments of inertia $I_1$ and $I_2$ 
\cite{mixings:2004}:
\begin{equation}
  \label{eq:massspl1}
\Delta M_{10-8}=\frac{3}{2}\frac{1}{I_{1}},\;\;\;
\Delta M_{\overline{10}-8}=\frac{3}{2}\frac{1}{I_{2}}.  
\end{equation}
The center of the baryon octet is just the average of the $\Lambda$
and $\Sigma$ masses, i.e. $M_{8}=1151.5\,\textrm{MeV}$, whereas the
center of the baryon decuplet is determined by
the $\Sigma^*$, i.e. $M_{10}=1385\,\textrm{MeV}$~\cite{PDG:2006}. 
In general, the baryon octet must satisfy the Gell-Mann-Okubo mass
formula~\cite{octet_relations}:
\begin{eqnarray} 
2(m_{N}+m_{\Xi}) & = & 3m_{\Lambda}+m_{\Sigma}.
\label{Gell-Mann-Okubo}
\end{eqnarray}
In the $\chi$QSM, we obtain for the baryon decuplet and anti-decuplet
the equal-spacing mass formulae as follows:
\begin{eqnarray}
m_{\Sigma^{*}}-m_{\Delta} & =m_{\Xi^{*}}-m_{\Sigma^{*}}= &
m_{\Omega}-m_{\Xi^{*}},\cr m_{N_{\overline{10}}}-m_{\Theta} &
=m_{\Sigma_{\overline{10}}}-m_{N_{\overline{10}}}= &
m_{\Xi_{\overline{10}}}-m_{\Sigma_{\overline{10}}}.
\end{eqnarray}
Since, in the $\chi$QSM, all baryons emerge from one classical
configuration, we also have the Guadagnini relation that connects 
baryon masses of the octet with those of the
decuplet~\cite{Guadagnini:1984}: 
\begin{eqnarray}
8(m_{\Xi^{*}}+m_{N})+3m_{\Sigma} & = & 11 m_{\Lambda}+8m_{\Sigma^{*}}.
\label{Guadagnini}
\end{eqnarray}
Using the numerical results listed in Table~\ref{Numerical-values} and
wave functions in Eq.(\ref{wavefunctions}), we obtain the mass
differences within a multiplet: 
\begin{eqnarray} 
\Delta M_{B_{1}B_{2}} & = & M_{B_{1}}-M_{B_{2}}=\langle
B_{1}|H_{\textrm{sb}}|B_{1}\rangle-\langle
B_{2}|H_{\textrm{sb}}|B_{2}\rangle.
\end{eqnarray} 
\begin{table}[h]
\caption{\label{Table2} Mass splittings between the baryon octet and
the baryon decuplet and anti-decuplet with three different values of
the constituent quark mass $M$.  The preferred value is $M=420$MeV.} 
\begin{center}
\begin{tabular}{c|cccc}
\hline
$M[\textrm{MeV}]$&
$400$&
$420$&
$450$&
exp.\tabularnewline
\hline
$\Delta M_{10-8}$&
$257$&
$279$&
$308$&
$234$\tabularnewline
$\Delta M_{\overline{10}-8}$ &
$558$&
$617$&
$673$&
-\tabularnewline
\hline
\end{tabular}
\end{center}
\end{table}
In Table~\ref{Table2} the results for the mass splittings are listed,  
where the values with $M=420$ MeV are the relevant ones for all 
applications of the $\chi$QSM.  The octet-decuplet
splitting deviates from the experimental data by about
$52\,\textrm{MeV}$.  

\begin{table}[h]
\caption{\label{table3}The baryon octet and decuplet mass splittings
in the $\chi$QSM given in MeV for the constituent
quark mass of $M=420$MeV. }
\centering
\begin{tabular}{c|cccc|ccc}
\hline
&
$\Lambda-N$&
$\Sigma-N$&
$\Sigma-\Lambda$&
$\Xi-\Sigma$&
$\Sigma_{10}^{*}-\Delta$&
$\Xi_{10}^{*}-\Sigma_{10}^{*}$&
$\Omega-\Xi_{10}^{*}$\tabularnewline
\hline
$\langle B|H_{\textrm{sb}}|B\rangle$&
$123$&
$177$&
$55$&
$96$&
$103$&
$103$&
$103$\tabularnewline
exp.&
$175$&
$250$&
$75$&
$124$&
$155$&
$145$&
$142$\tabularnewline
\hline
\end{tabular}
\end{table}
In Table~\ref{table3} the results of the mass splittings in the baryon
octet and decuplet are listed.  They are obtained by calculating the
matrix elements of the symmetry-breaking part of the collective
Hamiltonian in Eq.(\ref{eq:Ham}).        
Generally, also for other constituent quark masses, the results of the
$\chi$QSM with the parameter given in Table~\ref{Numerical-values}
underestimate the mass splittings for the hyperons by up to
$73\,\textrm{MeV}$.  Note that the deviation to the experimental data
in this work are larger than that in Ref.~\cite{Christov:1995vm}.  It
is due to the facts that in the present work we do not consider the 
quadratic $m_{s}$ corrections for the Hamiltonian and,
moreover, different numerical settings such as the size of the box for
solving the Dirac equation and regularization parameters end up with
slightly different results such as the $\Sigma_{\pi N}$ term: For
example, in Ref.~\cite{Blotz_mass_splittings} it is obtained to be
$56.14\,\textrm{MeV}$ while in the present work we get $\Sigma_{\pi
  N}=41\,\textrm{MeV}$.  In particular, the $\Sigma_{\pi N}$ term is
rather sensitive to the scheme of the regularization because of its
quadratic divergence.\\
Turning now to the anti-decuplet. We can calculate the masses and
splittings of the anti-decuplet baryons by
\begin{equation}
M_{B_{\overline{10}}}=M_8^{\mathrm{exp}} +
\Delta M_{\overline{10}-8}^{\chi\mathrm{QSM}} +
\Delta M_{\Delta^{\overline{10}}}^{\chi\mathrm{QSM}}. 
\end{equation}
The hypercharge splittings of the anti-decuplet are listed in
Table~\ref{AD masses}.  The experimental data are taken from the 
GRAAL experiment \cite{Kuznetsov:2004gy} for the $N_{\overline{10}}^{*}$
and from the NA49 experiment \cite{NA49_XI} for
$\Xi_{\overline{10}}^{*}$, though the NA49 data is still controversial.
\begin{table}[h]
\caption{\label{AD masses} The mass splittings of the baryon
anti-decuplet in the self-consistent $\chi$QSM in unit of MeV and for
the constituent quark mass $M=420$MeV.} 
\centering
\begin{tabular}{c|cccc}
\hline
&
$N_{\overline{10}}-\Theta$&
$\Sigma_{\overline{10}}-N_{\overline{10}}$&
$\Xi_{\overline{10}}-\Sigma_{\overline{10}}$&
$\Xi_{\overline{10}}-\Theta$\tabularnewline
\hline
$\langle B|H_{\textrm{sb}}|B\rangle$&
$115$&
$115$&
$115$&
$345$\tabularnewline
exp.&
$135$&
$ $&
$ $&
$322$\tabularnewline
\hline
\end{tabular}
\end{table}
\begin{table}[h]
\caption{\label{table6} Masses of the baryon octet, of the decuplet,
and of the anti-decuplet in unit of MeV.  We started from the experimental octet
center $M_{8}=1151.5$ and used the $\chi$QSM mass-splittings for
$M=420$MeV. Experimental values are written in parentheses and are taken
from the PDG \cite{PDG:2006}, from the GRAAL~\cite{Kuznetsov:2004gy}
and from the NA49 collaboration~\cite{NA49_XI}.} 
\begin{center}
\begin{tabular}{cccc}
\hline
&
Octet&
Decuplet&
Anti-decuplet\tabularnewline
\hline
$\Theta$&
$-$&
$-$&
$1538\,(1540)$\tabularnewline
$N/\Delta$&
$1001\,(939)$&
$1329\,(1232)$&
$1653\,(1675)$\tabularnewline
$\Lambda/\Sigma$&
$\Big{\{} \begin{array}{c}
1124 \, (1116)\\
1179 \, (1189)\end{array} $&
$1431\,(1385)$&
$1768\,(-)$\tabularnewline
$\Xi$&
$1275\,(1318)$&
$1533\,(1530)$&
$1883\,(1862)$\tabularnewline
$\Omega$&
$-$&
$1635\,(1672)$&
$-$\tabularnewline
\hline
\end{tabular}
\end{center}
\end{table}
With the set of parameters given in Table~\ref{Numerical-values},  
we obtain the results for the baryon masses in a qualitative agreement    
with the data.\\
With the $\chi$QSM values given in Tab.~\ref{table6} we calculate also
the Gell-Mann-Okubo and Guadagnini relations. Even though the absolute
masses of the $\chi$QSM are off by $(8-97)$ MeV both relations are well
satisfied, Table~\ref{table4}.

\begin{table}[h]
\caption{\label{table4}The Gell-Mann-Okubo and Guadagnini relations in
  the self-consistent $\chi$QSM given in unit of GeV and for the
  constituent quark mass $M=420$MeV.}
\centering
\begin{tabular}{c|c|c|c|c}
\hline
$[\mathrm{GeV}] $&
$2N+2\Xi$&
$3\Lambda+\Sigma$&
$8(m_{\Xi^{*}}+m_{N})+3m_{\Sigma}$&
$11m_{\Lambda}+8m_{\Sigma^{*}}$\tabularnewline
\hline
$\chi$QSM&
$4.55$&
$4.55$&
$23.81$&
$23.81$\tabularnewline
experiment&
$4.51$&
$4.54$&
$23.32$&
$23.36$\tabularnewline
\hline
\end{tabular}
\end{table}
  
\subsection{Axial-Vector Transition Constants }
We first present the results for the nonstrangeness ($\Delta S=0$)
$B_{10}\to B_{8}$ transitions, using Eq.(\ref{eq:delta-axial}) with
$a=\lambda^{3}/2$.  In Table~\ref{tab:deaxial} we list the $\Delta
S=0$ axial-vector transition constants $C_A^*$ for the $B_{10}\to
B_{8}$.  Since the leading ($\Omega^{0},m_{s}^{0}$)-order and the
$1/N_{c}$ rotational corrections ($\Omega^{1},m_{s}^{0}$) in
Eq.(\ref{eq:delta-axial}) do always {\it constructively} interfere,
the effects of the $m_{s}$-corrections turn out to be rather small,
i.e. they contribute to the axial-vector transition constants by about
$7$\%.  
\begin{table}[h]
\caption{\label{tab:deaxial}Axial-vector transition constants for the
$B_{10}\to B_{8}+\pi^0$  processes with $\Delta S=0$ and using
$M=420$ MeV.  In the last line the final results are listed with the
SU(3) symmetry breaking included.} 
\begin{center}
\begin{tabular}{c|cccc}
\hline
$C_{A}^{*}$&
$\Delta^{+}\to p$&
$\Sigma_{10}^{0}\to\Lambda$&
$\Sigma_{10}^{+}\to\Sigma^{+}$&
$\Xi_{10}^{0}\to\Xi^{0}$\tabularnewline
\hline
$m_{s}^{0}$&
$-0.89$&
$-0.77$&
$0.45$&
$0.45$\tabularnewline
$m_{s}^{0}+m_{s}^{1}$&
$-0.96$&
$-0.82$&
$0.45$&
$0.46$\tabularnewline
\hline
\end{tabular}
\end{center}
\end{table}

We will present now the results of the anti-decuplet transitions
results.  The anti-decuplet axial-vector transition constants
$g_{A}^{*}$ for $B_{\overline{10}}\to B_{8}+m$ are listed in
Table~\ref{tab:ac}.  In the case of pion- and eta-transitions, the
corresponding operators in Eq.(\ref{eq:axial-model}) are with
$a=\lambda^{3}$ and $a=\lambda^{8}$, respectively, and for kaon
transitions $a=\frac{1}{2}(\lambda^{4}\pm i\lambda^{5})$.  We find
that the leading-order contributions ($\Omega^{0},m_{\mathrm{s}}^{0}$)
and the rotational corrections ($\Omega^{1},m_{\mathrm{s}}^{0}$)
interfere always { \it destructively}, so that the axial-vector
transition constants turn out to be rather small in the chiral limit.
Due to this cancellation, the $m_{\mathrm{s}}$ corrections become
relevant.  As for the $\Theta^{+}\to n$ transitions, the
$m_{\mathrm{s}}$ corrections reduce even further $g_{A}^{*(\Theta\to
  n)}$ by about $40\,\%$ still being corrections in the
present formalism
\begin{equation}
g^*_A(\Theta nK): \,\,\,
\frac{|g^{*(m^1_s)}_A|}{|g^{*(\Omega^0,m^0_s)}_A| +
  |g^{*(\Omega^1,m^0_s)}_A| + |g^{*(m^1_s)}_A|} = \frac{0.036}{|0.31|
  + |-0.22|+|0.036|}=0.06.
\end{equation}
The mixing of the octet wave functions in the
$p_{\overline{10}}$ transitions are, compared to other transitions,
large. The total $m_{\mathrm{s}}$ corrections reduce
$g_{A}^{*(p_{\overline{10}}\to n \pi)}$ but increase
$g_{A}^{*(p_{\overline{10}}\to\Lambda K)}$ and
$g_{A}^{*(p_{\overline{10}}\to p \eta)}$.  However, for
$g_{A}^{*(p_{\overline{10}}\to n \pi)}$, various parts cancel 
each other, sometimes almost completely.  
\begin{table}[h]
\caption{\label{tab:ac}Axial-vector transition constants for
the $B_{\overline{10}}\to B_8$ using the self-consistent $\chi$QSM
with $M=420$MeV. Each column shows each contribution. 
The $m_{\mathrm{s}}$ corrections are listed separately: The
wave-function corrections from the mixing with the octet, those from
the 27-plet mixing, and the operator corrections, respectively.
}
\begin{center}
\begin{tabular}{c|r|r|r|r|r|r}
\hline
$g_{A}^{*}$&
$\Omega^{0},m_{s}^{0}$&
$\Omega^{1},m_{s}^{0}$&
$m_{s}^{1},\textrm{wf(8)}$&
$m_{s}^{1},\textrm{wf(27)}$&
$m_{s}^{1},\textrm{op}$&
total\tabularnewline
\hline
$\Theta^{+}\to n K^+$&
$0.310$&
$-0.220$&
$-0.013$&
$-0.024$&
$0.001$&
$0.053$\tabularnewline
$p_{\overline{10}}\to n \pi^+$&
$0.180$&
$-0.130$&
$-0.102$&
$0.033$&
$0.002$&
$-0.017$\tabularnewline
$p_{\overline{10}}\to\Lambda K^+$&
$-0.220$&
$0.160$&
$-0.068$&
$0.034$&
$-0.004$&
$-0.098$\tabularnewline
$p_{\overline{10}}\to p \eta$&
$-0.310$&
$0.220$&
$-0.042$&
$0.024$&
$0.0003$&
$-0.107$\tabularnewline
$p_{\overline{10}} \to \Delta^+ \pi^0$&
$0$&
$0$&
$0.074$&
$-0.154$&
$0.005$&
$-0.075$\tabularnewline
$\Xi_{\overline{10}}^{+}\to\Xi^{0} \pi^+$&
$-0.310$&
$0.220$&
$0$&
$-0.016$&
$-0.014$&
$-0.120$\tabularnewline
$\Xi_{\overline{10}}^{--}\to\Sigma^{-} K^-$&
$-0.310$&
$0.220$&
$-0.013$&
$-0.008$&
$-0.007$&
$-0.118$\tabularnewline
\hline
\end{tabular}
\end{center}
\end{table}

One should note that in SU(3) flavor symmetry, the transitions from
the baryon anti-decuplet to the decuplet are strictly forbidden,
because the direct product of the baryon antidecuplet and the octet
current does not contain the decuplet in its irreducible
representation.  However, if we turn on the $m_{\mathrm{s}}$
corrections, the transitions from the anti-decuplet to the decuplet
are allowed, since the anti-decuplet mixes with the octet and 27-plet
according to Eq.(\ref{wavefunctions}).  We will concentrate in the
present work on the $N_{\overline{10}}\to\Delta$ transition only.
In the last line in Table~\ref{tab:ac} each contribution to the
axial-vector transition constant for the
$p_{\overline{10}}\to\Delta^+$ process is listed.  As shown in the
last row of Table~\ref{tab:ac}, there is no contribution from the
leading and rotational $1/N_c$ orders, but we get small contributions
from the $m_{\mathrm{s}}$ corrections.  It is found that the 
wave-function correction due to the 27-plet mixing turns out to be
large and has the opposite sign to the octet mixing one. The
$m_{\mathrm{s}}$ corrections from the operators are negligible.
\subsection{Decay Widths}
Now, we are in a position to calculate the decay widths for the
transitions between different baryons, using Eq.(\ref{decay_formula}).
We first consider the nonstrangeness transitions from the baryon
decuplet to the octet.  In Table~\ref{decdecay}, we list the
corresponding results with $M=420$ MeV.
\begin{table}[h]
\caption{\label{decdecay}Decay widths of the $\Delta S=0$ transitions 
from the baryon decuplet to the octet with $M=420\,\textrm{MeV}$.} 
\begin{center}
\begin{tabular}{c|cccc}
\hline
&
$\Delta^{+}\to p\pi^{0}$&
$\Sigma^{*0}\to\Lambda\pi^{0}$&
$\Sigma^{*+}\to\Sigma^{+}\pi^{0}$&
$\Xi^{*0}\to\Xi^{0}\pi^{0}$\tabularnewline
\hline
$\Gamma\,[\textrm{MeV}]$&
$48.7$&
$30.3$&
$2.2$&
$3.9$\tabularnewline
\hline
\end{tabular}
\end{center}
\end{table}
We have assumed here isospin symmetry. Calculating relative isospin
factors, we can evaluate the total decay width for each channel.
Summing all possible transitions and averaging over 
the initial states, we obtain
\begin{eqnarray}
\Gamma(\Delta\to N\pi)& = & \frac{3}{2} \Gamma(\Delta^{+}\to
p\pi^{0}),\;\;\;\;\;\; \Gamma(\Sigma^* \to \Lambda\pi) =
\Gamma(\Sigma^{*+} \to \Lambda\pi^{0}),\cr 
\Gamma(\Sigma^* \to \Sigma\pi) & = & 2 \Gamma(\Sigma^{*+}
\to \Sigma^{+}\pi^{0}),\;\;\;  \Gamma(\Xi^* \to \Xi\pi^{0}) = 2
 \Gamma(\Xi^{*0} \to \Xi^{0}\pi^{0}).
\end{eqnarray}
The total decay widths for $\Delta S=0$ transitions are listed in
Table~\ref{dectow}. 
\begin{table}[h]
\caption{\label{dectow}Total decay widths for the decuplet to octet
transitions with  $\Delta S=0$ using self-consistent $\chi$QSM
with $M=420\,\textrm{MeV}$. For the $\Xi^{*}$ transitions, we take the
experimental data from Ref.~\cite{PDG:2006}: 1)
$\Gamma(\Xi^{-}\to\Xi^{0})+\Gamma(\Xi^{-}\to\Xi^{-})$ 
and 2) $\Gamma(\Xi^{0}\to\Xi^{-})$. }
\begin{center}
\begin{tabular}{c|cccc}
\hline
&
$\Delta\to N\pi$&
$\Sigma^* \to\Lambda\pi$&
$\Sigma^* \to\Sigma\pi$&
$\Xi_{10}\to\Xi\pi$\tabularnewline
\hline
$\Gamma^{\chi\textrm{QSM}}\,[\textrm{MeV}]$&
$73.1$&
$30.3$&
$4.4$&
$7.9$\tabularnewline
$\Gamma^{\textrm{PDG}}\,[\textrm{MeV}]$&
$111.5$&
$36.1$&
$ $&
$16.2^{1)}(11.0^{2)})$\tabularnewline
\hline
\end{tabular}
\end{center}
\end{table}

Even though the values of the present work underestimates the
decuplet widths they are comparable to the decuplet widths given in
Ref.~\cite{Diakonov:1997mm} by taking the presented formulae
literally. A clarification of the given numbers in
Ref.~\cite{Diakonov:1997mm} can be found in
Ref.~\cite{DPP_comment_on_jaffe}. 

It is also of great interest to compare the ratio of the decay widths
for $\Sigma^*\to \Sigma$ and for $\Sigma^*\to \Lambda$.  In the
present work, we obtain the ratio as follows:
\begin{equation}
  \label{eq:ratio3}
\frac{\Gamma(\Sigma^* \to \Sigma)}{\Gamma(\Sigma^* \to \Lambda)} =
0.145,   
\end{equation}
which is in good agreement with the data from the particle data
group~\cite{PDG:2006}: $\Gamma(\Sigma^* \to \Sigma) / \Gamma(\Sigma^*
\to \Lambda) = 0.135\pm0.011$.

We now consider the decays from the baryon anti-decuplet to the
octet.  Based on the results of the axial-vector transition constants
listed in Table~\ref{tab:ac}, we can immediately calculate the decay
widths for the $B_{\overline{10}} \to B_8$ transitions.  The results are
presented in Table~\ref{decay-width decomposed}.
\begin{table}[h]
\caption{\label{decay-width decomposed}Partial decay widths for
the $B_{\overline{10}}\to B_8$ transitions in the self-consistent
$\chi$QSM using $M=420$ MeV. 
The $m_{\mathrm{s}}$ corrections from the operators 
are added to yield the total results.}
\begin{center}
\begin{tabular}{c|c|c|c|c|c}
\hline
$\Gamma\,[\textrm{MeV}]$&
$\Omega^{0}$&
$\Omega^{0}+\Omega^{1}$&
$\Omega^{0}+\Omega^{1}+\textrm{wf(8)}$&
$\Omega^{0}+\Omega^{1}+\textrm{wf(8+27)}$&
total\tabularnewline
\hline
$\Theta^{+}\to nK^{+}$&
$12.23$&
$1.04$&
$0.77$&
$0.36$&
$0.36$\tabularnewline
$p_{\overline{10}}\to n\pi^{+}$&
$53.38$&
$4.11$&
$4.45$&
$0.59$&
$0.47$\tabularnewline
$p_{\overline{10}}\to\Lambda K^{+}$&
$3.16$&
$0.23$&
$1.07$&
$0.58$&
$0.63$\tabularnewline
$p_{\overline{10}}\to p\eta$&
$16.92$&
$1.43$&
$3.07$&
$2.05$&
$2.04$\tabularnewline
$p_{\overline{10}} \to \Delta^+ \pi^0$&
$0$&
$0$&
$4.52$&
$5.28$&
$4.64$\tabularnewline
$\Xi_{\overline{10}}^{+}\to\Xi^{0}\pi^{+}$&
$80.33$&
$6.77$&
$6.77$&
$9.39$&
$12.03$\tabularnewline
$\Xi_{\overline{10}}^{--}\to\Sigma^{-}K^{-}$&
$31.32$&
$2.64$&
$3.46$&
$4.02$&
$4.54$\tabularnewline
\hline
\end{tabular}
\end{center}
\end{table}
Since the axial-vector transition constants for the
$B_{\overline{10}}\to B_8$ turn out to be rather small, we get
consequently the small decay widths for the baryon anti-decuplet to
the octet transitions.  

Evaluating relative isospin factors, we get from these transitions 
the total decay widths. Summing all transitions and averaging
over initial states, we obtain
\begin{eqnarray}
\Gamma(\Theta\to NK) & = & 2 \Gamma(\Theta\to nK^{+}),\;\;\;\;\;\;\; 
\Gamma(N_{\overline{10}}\to
N\pi) = \frac{3}{2}\Gamma(p_{\overline{10}}\to n\pi^{+}),\cr 
\Gamma(N_{\overline{10}}\to\Lambda
K) & = & \Gamma(p_{\overline{10}} \to \Lambda K^{+}),\;\;\;\;\;\;
\Gamma(N_{\overline{10}}\to N\eta) = \Gamma(p_{\overline{10}}\to
p\eta),\cr
\Gamma(\Xi_{\overline{10}}\to\Xi\pi) & = &
\Gamma(\Xi_{\overline{10}}^{+}\to\Xi^{0}\pi^{+}),\;\;\;\;\; 
\Gamma(\Xi_{\overline{10}}\to\Sigma
K) = \Gamma(\Xi_{\overline{10}}^{--}\to\Sigma^{-}K^{-}),
\end{eqnarray}
and for the $N_{\overline{10}}\to\Delta$ transitions
\begin{equation} 
\Gamma(N\to\Delta\pi)  = 
3\times\Gamma(p_{\overline{10}}\to\Delta^{+}\pi^{0})\,\,.
\end{equation}
The total decay widths are presented in Table~\ref{tab:totald}.  
\begin{table}[h]
\caption{\label{tab:totald} Final result for the total decay widths for the
  $B_{\overline{10}}\to B_8$ transitions in unit of MeV, as varying
  $M$ from 400 to 450 MeV in the self-consistent $\chi$QSM. The results
for $M=420$MeV are our prefered values.}
\begin{center}
\begin{tabular}{c|ccccccc}
\hline
$M\,[\textrm{MeV}]$&
$\Theta\to NK$&
$N_{\overline{10}}\to N\pi$&
$N_{\overline{10}}\to\Lambda K$&
$N_{\overline{10}}\to N\eta$&
$N_{\overline{10}} \to \Delta \pi $&
$\Xi_{\overline{10}}\to\Xi_{8}\pi$&
$\Xi_{\overline{10}}\to\Sigma_{8}K$\tabularnewline
\hline
$400$&
$0.46$&
$1.54$&
$0.63$&
$1.85$&
$17.06$&
$11.60$&
$4.23$\tabularnewline
$420$&
$0.71$&
$0.71$&
$0.63$&
$2.04$&
$13.92$&
$12.03$&
$4.54$\tabularnewline
$450$&
$1.01$&
$0.09$&
$0.63$&
$2.09$&
$11.45$&
$12.24$&
$4.75$\tabularnewline
\hline
\end{tabular}
\end{center}
\end{table}
In Ref.~\cite{Diakonov:1997mm,jaffe_on_org_paper,DPP_comment_on_jaffe}
the formulae for the decay width of the $\Delta\to N\pi$ and $\Theta
\to NK$ transitions are given as  
\begin{eqnarray}
\Gamma(\Delta\to N \pi ) &=&
\frac{3(G_{0}+\frac{1}{2}G_{1})^{2}}{2\pi(M_{\Delta}+M_{N})^{2}}|{\bm 
  k}_{\pi}|^{3}\frac{M_{N}}{M_{\Delta}}\frac{1}{5} , \cr
\Gamma(\Theta \to N K ) &=& \frac{3(G_{0}-G_{1})^{2}}{2\pi(M_{\Theta}
  + M_N)^{2} }|{\bm
  k}_{K}|^{3}\frac{M_{N}}{M_{\Theta}}\frac{1}{5}, \label{eq:gammad} 
\end{eqnarray}
and for the $g_{\pi NN}$ constant in the $\chi$QSM as
\begin{equation} 
g_{\pi NN} = \frac{7}{10}(G_{0} + \frac{1}{2}G_{1}),
\label{eq:gpNN}
\end{equation} 
where terms proportional to $G_2$ and $c_{\overline{10}}$ were dropped.
In Ref.~\cite{Diakonov:1997mm}, the coupling constant of
$G_{0}+\frac{1}{2}G_{1}=19$ is used, which follows from inverting 
Eq.(\ref{eq:gpNN}) with the experimental value $g_{\pi NN}=13.6$.  In
order to separate $G_{0}$ and $G_{1}$, Ref.~\cite{Diakonov:1997mm} has
used the parameter $G_{1}/G_{0}=0.4$.  We will comment on this ratio
later in detail.  In Ref.~\cite{jaffe_on_org_paper}, $\Gamma(\Delta N
\pi)$ in Eq.(\ref{eq:gammad}) is inverted by using the experimental
value $\Gamma(\Delta \to N\pi)=110$MeV in order to obtain
$G_{0}+\frac{1}{2}G_{1}=25$ which would give a large $g_{\pi
  NN}=17.5$.  Reference~\cite{jaffe_on_org_paper} claimed that the 
decay widths should be $\Gamma(\Delta \to N\pi)\approx 68$MeV and
$\Gamma(\Theta \to NK) \leq 30$ MeV compared to the in
Ref.~\cite{Diakonov:1997mm} given values of $\Gamma(\Delta \to
N\pi)\approx 110$ MeV and $\Gamma(\Theta \to NK) \leq 15$ MeV. 
Reference~\cite{DPP_comment_on_jaffe} clarifies the situation.
Furthermore, in Ref.~\cite{DPP_comment_on_jaffe}, the authors emphasized 
that the value of $\Gamma(\Theta N K) \leq 15$MeV is the most
conservative prediction and that by changing the ratio $G_1/G_0$ from
$0.4-0.6$ the decay width varies between $(11.2-3.6)$MeV.

The ratio $G_1/G_0$ is the only input depending of a certain model and
originally this ratio was taken from the $\chi$QSM calculations
Refs.~\cite{Christov:1993ny,Blotz:1994wi}. The present work is based
on the same formalism as used and developed in
Refs.~\cite{Christov:1993ny,Blotz:1994wi}, however, several parts have
been optimized since then. The symmetry conserving quantization
\cite{Praszalowicz:1998jm} was established after the publication of
Ref.\cite{Diakonov:1997mm} and has been applied since then for all
octet baryon obsevables within the present formalism. The ratio
$G_1/G_0$ corresponds to the ratio $a_2/a_1$ of the present work, 
where $a_i$ are defined in Eq.(\ref{eq:as}). In the present work this
ratio is $a_2/a_1=-0.68$, i.e. $G_1/G_0 = 0.68$.

The difference of these ratios lies in the fact that
Ref.~\cite{Blotz:1994wi} has employed the collective quantization
without symmetry conservation. The $a_{1}$ would increase
noticeably without symmetry-conserving quantization and therefore
the ratio $a_{2}/a_{1}$ would be lessened.  Since we utilize in the
present work the symmetry-conserving
quantization~\cite{Praszalowicz:1998jm}, we can correctly calculate
the parameters $a_i$.  Had Ref.\cite{Diakonov:1997mm} used the ratio
$G_{1}/G_{0}=0.68$,  the decay width in \cite{Diakonov:1997mm} would
have turned out to be $\Gamma(\Theta NK)=3.4$ MeV, which is much
smaller than the value $\Gamma(\Theta NK)<15$ MeV published in
Ref.\cite{Diakonov:1997mm}, whereas $\Gamma(\Delta N\pi)$ would remain
unchanged.  In addition, following the comment of
Ref.\cite{jaffe_on_org_paper} with $G_{1}/G_{0}=0.68$, one will end up
with the decay width of the $\Theta$ which is also smaller than
predicted in Ref.\cite{Diakonov:1997mm}. The predicted 
physics in Ref.\cite{Diakonov:1997mm} by using the $\chi$QSM is
therefore unchanged. At this point we also want to emphasize again the
fact that $\Gamma(\Theta K N)$ vanishes in the non-relativistic limit
as figured out in Ref.\cite{Diakonov:1997mm}. The whole
$\Theta$ decay width is a function of $G_0-G_1-G_2/2$ where the
non-relativistic quark model predicts $G_1 = G_0 \cdot 4/5$ and
$G_2 =G_0 \cdot 2/5$. In this context it is interesting that the
symmetry-conserving quantization is achieved by demanding the SU(3)
and SU(2) $\chi$QSM versions to have the same non-relativistic limit.

As shown in Appendix~\ref{app:bame}, the value of
$G_{0}+\frac{1}{2}G_{1}$ in the $\chi$QSM, i.e. in the present work,
turns out to be $13.43$ instead of $19$ which is used in
Ref.~\cite{Diakonov:1997mm}.  Then, the decay width $\Gamma(\Theta
NK)=3.4$MeV would be further reduced by about $50\,\%$, i.e. $\Gamma(\Theta
NK)=1.7$ MeV.  Note that Ref.~\cite{Diakonov:1997mm} has only
considered the wave-function corrections due to the mixing with the
octet, while in the present work we consider all possible mixings.  If
we keep only the mixing with the octet, then we obtain the decay width
$\Gamma(\Theta NK)=1.54$ MeV.  

The $\Theta NK$ coupling constants from the present results are
yielded as follows:
\begin{equation}
g_{\Theta NK}^{(m_{s}^{0})}=1.41,\;\;\; g_{\Theta
  NK}^{(m_{s}^{0}+m_{s}^{1})}=0.83
\end{equation}  
without and with SU(3) symmetry breaking effects taken into account,
respectively.   The corresponding decay widths for the $\Theta\to NK$ 
transition are then evaluated as 
\begin{center}
\begin{tabular}{cc}
\hline
\tabularnewline
$\Gamma^{(m_{s}^{0})}(\Theta
NK)=2.08\,\textrm{MeV},\;\;\;$&
$ \Gamma^{(m_{s}^{0}+m_{s}^{1})}(\Theta
NK)=0.71\,\textrm{MeV}\,\,\,.\label{eq:final1}$\tabularnewline
\tabularnewline
\hline
\end{tabular}
\end{center}
It is interesting to compare this final result of 
$\Gamma^{(m_{s}^{0}+m_{s}^{1})}(\Theta NK)=0.71\textrm{MeV}$ with
the experimental data of the DIANA collaboration $\Gamma(\Theta
NK)=0.36\pm0.11$ MeV~\cite{DIANA2}. The $\Delta / \Theta$
decay ratios are given in this work as follows:
\begin{equation} 
\frac{\Gamma(\Delta N \pi)}{\Gamma(\Theta N K)} \Big{|}_{ \chi
  \textrm{QSM} } = \frac{73.1\textrm{MeV}}{0.71\textrm{MeV}}=103
\,\,\,\,\,\, , \,\,\,\,\,\, \frac{\Gamma(\Delta N \pi)}{\Gamma(\Theta
  N K)}\Big{|}_{   \textrm{exp.} } =
\frac{111.5\textrm{MeV}}{0.36\textrm{MeV}}= 310 \,\,\,. 
\end{equation} 
The results of the $\chi$QSM are therefore compatible with the
smallness of the $\Theta$ decay width, compared to the $\Delta$ decay
width.  Considering the fact that in the present work we do not have
any adjustable free parameter except for the constituent quark mass
$M$ that is also fixed to be $420$ MeV~\footnote{One should 
note that the four parameters (cut-off mass $\Lambda$, current quark
mass $\overline{m}$, constituent quark mass $M$, and strange current
quark mass $m_{\mathrm{s}}$) of the effective chiral action in the
$\chi$QSM have been adjusted many years ago to $f_{\pi}$, $m_{\pi}$,
the proton charge radius, and SU(3) baryon mass splittings.},
it is remarkable for the present results to be in such agreement with 
the DIANA data.  

Projecting the $\chi$QSM-soliton upon its $3-$ and
$5-$quark components in the light-cone basis, a value of
$\Gamma(\Theta NK)=2.26$ MeV is yielded for the SU(3)-symmetric case
(i.e. $m_s=0$)~\cite{IMF_imp_width}.  Imposing the condition of the
energy-momentum conservation in the same method as in
Ref.~\cite{IMF_imp_width}, Ref.~\cite{Lorceprivcomm} has shown the 
decay width to be $\Gamma(\Theta NK)\sim 0.4$ MeV.  A
"model-independent approach" in the $\chi$QSM gives the decay width
$\Gamma^{(m_{s}^{0}+m_{s}^{1})}(\Theta NK)=0.76\textrm{MeV}$ with the
singlet axial-vector constant $g_{A}^{0}=0.36$ and $\Sigma_{\pi N}=45$ 
MeV~\cite{modelindep_theta_width}.  Note that these values are quite
similar to those computed in the $\chi$QSM with a self-consistent
soliton profile~\cite{SilvaKUG:2005}.  Thus all calculations
based on the $\chi$QSM produce a small value of the decay width,
i.e. $\Gamma(\Theta NK) \leq 1\,\textrm{MeV}$, which is consistent
with the recent DIANA measurement~\cite{DIANA2}.

In Ref.~\cite{Arndt:2003ga}, the decay widths of non-strange
partners of the $\Theta^{+}$ have been investigated, where the decay widths are found to be
\begin{equation}
\Gamma(N_{\overline{10}}\Lambda K)=0.70,\;\;\;
\Gamma(N_{\overline{10}}N\eta) = 1.80,\;\;\;
\Gamma(N_{\overline{10}}N\pi)  = 2.10,\;\;\;
\Gamma(N_{\overline{10}}\Delta\pi) = 2.80,
\end{equation}
in unit of MeV.  However, taking the values of the present work and
taking only the mixing with the baryon octet into account, we get
\begin{equation}
\Gamma(N_{\overline{10}}\Lambda K)=1.07,\;\;\;
\Gamma(N_{\overline{10}}N\eta)=3.07,\;\;\;
\Gamma(N_{\overline{10}}N\pi)=6.67,\;\;\;
\Gamma(N_{\overline{10}}\Delta\pi)= 13.92,
\end{equation}
while considering all $m_{\mathrm{s}}$ corrections, we obtain our 
final results as follows: 
\begin{equation}
  \label{eq:finalwidth1}
\Gamma(N_{\overline{10}}\Lambda K)=0.63,\;\;\;
\Gamma(N_{\overline{10}}N\eta)=2.04,\;\;\;
\Gamma(N_{\overline{10}}N\pi)=0.71,\;\;\;
\Gamma(N_{\overline{10}}\Delta\pi)= 13.92.   
\end{equation}
At this point we want to stress that the $N\eta$ channel is stronger
than the $N\pi$ channel.  It is found that the inclusion of the
27-plet mixing has a large influence on the
$\Gamma(N_{\overline{10}}N\pi)$ and
$\Gamma(N_{\overline{10}}\Delta\pi)$.  Therefore, these two
transitions are more sensitive to the multiplet-mixing angles than 
the other.  For the $N_{\overline{10}}\to N\pi$ transition, the
27-plet mixing contributes destructively to the axial-vector constant
with the combined flavor SU(3)-symmetric part and octet mixing
correction.  The 27-plet mixing correction, being 
small compared to the $\Omega^0$ order, turns out to be sizeable,
since the $\Omega^0+\Omega^1$ contributions and the octet mixing are
almost canceled.  In the case of the
$\Gamma(N_{\overline{10}}\Delta\pi)$, which is only finite with flavor
SU(3)-breaking effects, the correction of the 27-plet mixing changes
the sign of the axial-vector constant, though the decay width happens
to be the same as that without it.  When it comes to the
$\Gamma(N_{\overline{10}}\Delta\pi)$ transition, we find that the
decay formula used in this work is different from that in
Ref.~\cite{Arndt:2003ga}. In Eq.(\ref{decay_formula}) there is
the mass of the $\Delta$ in the denominator, which comes due to the
mass factor in Eq.(\ref{appendixdecayformula}), whereas in other
formulae there are the corresponding mass of the decaying particle in
the denominator.  The difference yields about 50\% for the 
$\Gamma(N_{\overline{10}}\Delta\pi)$ decay width.

In Ref.~\cite{non_strange_partner2} the same authors have argued that
a larger value of the mixing angle between the octet and the
anti-decuplet is more probable, which would increase the decay width
to be $\Gamma(N_{\overline{10}}\Delta\pi)=15$ MeV. Thus, in this case,
the final result of this work is compatible with that of
Ref.~\cite{non_strange_partner2}, though in the present work we 
take into account 27-plet mixing corrections and a different decay-width
formula.  The authors of \cite{non_strange_partner2}
suggested a nucleon state $N(1680)$ as a pentaquark and 
concluded that the decay channels $N_{\overline{10}} \to \Delta\pi$,
though being flavor-SU(3) forbidden, and $N_{\overline{10}} \to N
\eta$ are the largest one.  The anti-decuplet nulceon state in the
$\chi$QSM of this work has a mass of $M=1654$ MeV and the decay
channels $N_{\overline{10}} \to \Delta\pi$ and $N_{\overline{10}} \to N
\eta$ are also more noticeable than the other.
\section{Summary and conclusion}
In the present work, we investigated the mass-splittings and strong
decays of the baryon anti-decuplet within the framework of the 
self-consistent SU(3) chiral quark-soliton model ($\chi$QSM).  We took linear
rotational $1/N_{c}$ and linear $m_{\mathrm{s}}$-corrections into
account and employed the symmetry conserving quantization which is
crucial in producing the small width of the $\Theta^+$.  All 
parameters used in the present work have been already fixed in
reproducing the pion and nucleon properties.  The general formalism
of this work corresponds to that in the $\chi$QSM done for many years, 
having been successfully applied to the baryon octet and decuplet
regime since the development and publication of the symmetry
conserving quantization.  No additional parameter 
has been adjusted in the calculation for the baryon anti-decuplet. We
used in the present work the self-consistent soliton profile in order
to solve numerically single-particle Dirac Hamiltonian in the chiral
quark-soliton model for the corresponding eigenvalues and eigentates
that are used in order to compute all observables.

Having considered the $1/N_c$ rotational corrections, we were able to
calculate the centered mass differences between the baryon octet and
the decuplet and anti-decuplet.  In addition, having taken strange
current quark masses into account, we also calculated the mass
splittings within a baryon multiplet.  We computed the absolute masses
of the baryon anti-decuplet in the $\chi$QSM,  
starting from the experimental octet center.  The final results of
this work are givn in Table~\ref{table6}.  Although this work
does not reproduce well the experimental octet and decuplet splittings, 
compared to previous $\chi$QSM works in which quadratic
$m_{\mathrm{s}}$ corrections were considered, the results are still in
qualitatively good agreement with experimental data. 

We also calculated the axial-vector constants for the baryon decuplet
to octet transitions, see Table~\ref{tab:deaxial}.  Applying the
generalized Goldberger-Treimann relation those constants are used to
calculate the strong couplings and evaluated the corresponding decay
widths, see Table~\ref{dectow} for the final results.  All decuplet
to octet decays obtained in the $\chi$QSM are in agreement with
earlier published results in the chiral solitonic picture 
and agree qualitatively with experimental data given by the particle
data group~\cite{PDG:2006}. 

Finally we applied the same techniques to the decays of the
baryon anti-decuplet, see Table~\ref{tab:ac} and
Table~\ref{tab:totald} for the final results.  In general, we found
that  those terms of the axial-vector transition constants associated
with the $\overline{10} \rightarrow 8$ decays are very small.  This
occurs due to a numerical cancelation of the leading contributions 
$(\Omega^{0},m_{\mathrm{s}}^{0})$ with the rotational corrections
$(\Omega^{1},m_{\mathrm{s}}^{0})$.  Those terms are constructively
interfering in the $10 \to 8$ transitions, while destructively
interfering in the $\overline{10} \to 8$ transitions.  Therefore,
the axial-vector constants for the $10 \to 8$ transitions become
large, while those for the $\overline{10} \to 8$ ones turn out to be
small.  These results immediately give the decay widths:   
$\Gamma\sim 70\,\mathrm{MeV}$ for $\Delta\to N\pi$ and
$\Gamma\leq 2\,\mathrm{MeV}$ for $N_{\overline{10}}\to N\pi$.  The
decay width of the $\Theta^{+}\to NK$ transition is found to be 
$\Gamma=0.71\,\mathrm{MeV}$ which is comparable to the latest DIANA
result: $\Gamma=0.36\pm0.11\,\mathrm{MeV}$. Because of the fact that 
the leading contributions are almost canceled with the subleading
rotational corrections, the $m_{\mathrm{s}}$ corrections give sizeable
effects on the axial-vector transition constants for the
$\overline{10}\to 8$, so that SU(3) symmetry-breaking effects turn out
to be about $50\,\%$ for the decay widths from the baryon antidecuplet
to the octet.  
We also investigated the SU(3)-forbidden decays from the
baryon anti-decuplet to the decuplet.  The transitions from the
anti-decuplet to the decuplet are entirely due to $m_{\mathrm{s}}$
corrections.  However, it turns out that the forbidden decay width for
$N_{\overline{10}}\to\Delta\pi$ and the width for the SU(3) allowed
channel $N_{\overline{10}}\to N \eta$ are larger than 
that for the decay $N_{\overline{10}}\to N\pi$, respectively.
\section*{Acknowledgments}
The authors are very grateful to J.K. Ahn, A. Hosaka, C. Lorce,
T. Nakano, S.i. Nam, M.V. Polyakov, M. Prasza\l owicz, and Gh.-S Yang
for helpful discussions and invaluable comments.  The work is
supported by the Transregio-Sonderforschungsbereich
Bonn-Bochum-Giessen, the Verbundforschung (Hadrons and Nuclei) of the
Federal Ministry for Education and Research (BMBF) of Germany, the
Graduiertenkolleg Bochum-Dortmund, the COSY-project J\"ulich as well
as the EU Integrated Infrastructure Initiative Hadron Physics Project  
under contract number RII3-CT-2004-506078.
The present work is also supported by the Korea Research Foundation
Grant funded by the Korean Government(MOEHRD) (KRF-2006-312-C00507).  
T. Ledwig was also supported by a DAAD doctoral exchange-scholarship.  

\begin{appendix}
\section{Decay forumlae \label{sec:Decay-forumlae}}
Using the explicit expression for the Dirac and Rarita-Schwinger
spinors and notations of \cite{PeskinSchroeder}
\begin{eqnarray}
\sum_{i}^{4}u_{\mu}^{i}(p)\overline{u}_{\nu}^{i}(p) &=& ( p^{\mu}
\gamma_{\mu} + m) \Big[-g_{\mu\nu} +
\frac{1}{3}\,\gamma_{\mu}\gamma_{\nu} +
\frac{1}{3m}(\gamma_{\mu}p_{\nu} - \gamma_{\nu}p_{\mu}) +
\frac{2}{3m^{2}} p_{\mu}p_{\nu}\Big],
\end{eqnarray}
we get the following invariant matrix elements 
\begin{eqnarray}
\sum_{s,s^{\prime}}|\mathcal{M}_{B_{\overline{10}}B_{8}m}|^{2} & = &
g_{B_{\overline{10}}B_{8}m}^{2}\,\,
2\Big((M_{\overline{10}}-M_{8})^{2}-m^{2} \Big)\\
\sum_{s,s^{\prime}}|\mathcal{M}_{\Delta N\pi}|^{2} & = & 
g_{\Delta N\pi}^{2} \,\,\frac{4}{3}\,\,\Big((m_{N} + m_{\Delta})^{2} -
m_{\pi}^{2}\Big){\bm k}_{\pi}^{2}\\
\sum_{s,s^{\prime}}|\mathcal{M}_{N^{*}\Delta\pi}|^{2} & = &
g_{N^{*}\Delta\pi}^{2}\,\,\frac{4}{3}\frac{M_{N^{*}}^{2}}{
M_{\Delta}^{2}}{\bm k}^{2}\Big((M_{N^{*}} + M_{\Delta})^{2} -
m_{\pi}^{2}\Big).
\label{appendixdecayformula}
\end{eqnarray}

The input masses for the mesons and baryons, and for the meson decay
constant are listed in Table~\ref{tab:app1} in unit of MeV.  
\begin{table}[h]
\caption{Input for the baryon and meson masses and for the meson
  decay constants.\label{tab:app1}}
  \centering
\begin{tabular}{|l|l|l|l|}
\hline
Octet & Decuplet & Anti-decuplet & Mesons \\
\hline
$M_{N}=939$&
$M_{\Delta}=1232$&
$M_{\Theta}=1540$&
$m_{\pi}=139$\tabularnewline
\hline
$M_{\Lambda}=1116$&
&
$M_{N_{\overline{10}}}=1675$&
$m_{K}=495$
\tabularnewline
\hline
$M_{\Sigma}=1189$&
$M_{\Sigma_{10}}=1385$&
&
$m_{\eta}=545$
\tabularnewline
\hline
$M_{\Xi}=1318$&
$M_{\Xi_{10}}=1530$&
$M_{\Xi_{\overline{10}}}=1862$&
$f_{\pi}=93$, $f_K=f_\eta=1.2f_\pi$ 
\tabularnewline
\hline
\end{tabular}
\end{table}

\section{Axial-vector densities in the $\chi$QSM \label{app:axdens} }
The densities $\mathcal{A}({\bm z}),\mathcal{B}({\bm z}),\cdots$ given
in the first part of Eq.(\ref{eq:axial-model})
for $\mathcal{G}_{A}^{\chi}({\bm z})$ have to be evaluated explicitly.
The corresponding densities in Eq.(\ref{eq: axial-densities thesis})
are given as follows: 
\begin{eqnarray}
\mathcal{A}({\bm z}) & = & N_c \langle v|{\bm z}\rangle\{
{\bm \sigma}_1\otimes{\bm\tau}_1\}_{0}\langle{\bm
  z}|v\rangle 
+ N_c\sum_{n}\sqrt{2G+1}\mathcal{R}_{1}(\varepsilon_{n})\langle 
n|{\bm z}\rangle\{ {\bm \sigma}_1\otimes{\bm\tau}_1\}_{0}\langle{\bm
  z}|n\rangle, \\
\mathcal{B}({\bm z}) & = & N_c
\sum_{\varepsilon_{n}\neq\varepsilon_{v}}
\frac{1}{\varepsilon_{v}-\varepsilon_{n}}(-)^{G_{n}}\langle 
n|{\bm z}\rangle {\bm \sigma}_1\langle{\bm z}|v\rangle\cdot\langle
v|{\bm\tau}_1|n\rangle \cr 
 &  & -\; \frac{N_c}{2}\sum_{n,m}(-)^{G_{n}-G_{m}}\langle m|{\bm
   z}\rangle {\bm \sigma}_1\langle{\bm z}|n\rangle\langle
 n|{\bm\tau}_1|m\rangle\mathcal{R}_{5}(\varepsilon_{n},
 \varepsilon_{m}),\\ 
\mathcal{C}({\bm z}) & = & N_c \sum_{n^{0}} \frac{1}{\varepsilon_{v} -
  \varepsilon_{n^{0}}} \langle v|{\bm z}\rangle\{ {\bm \sigma}_1
\otimes {\bm\tau}_1\}_{0}\langle{\bm z}|n^{0}\rangle\langle n^{0}| v
\rangle\cr
 &  & - \;N_c\sum_{n,m}\sqrt{2G_{n}+1}\langle n|{\bm z}\rangle\{
 {\bm \sigma}_1\otimes{\bm\tau}_1\}_{0}\langle{\bm z}|m^{0}\rangle\langle
 m^{0}|n\rangle\mathcal{R}_{5}(\varepsilon_{n},\varepsilon_{m^{0}}),\\
\mathcal{D}({\bm z}) & = & N_c
\sum_{n}\frac{\mathrm{sign}(\varepsilon_{n})}{\varepsilon_{v} -
  \varepsilon_{n}} (-)^{G_{n}}\langle
n|{\bm z}\rangle\{ {\bm \sigma}_1\otimes{\bm\tau}_1\}_{1}\langle{\bm
  z}|v\rangle\cdot\langle v|{\bm\tau}_1|n\rangle\cr 
 &  &  + \;  \frac{N_c}{2}\,\sum_{n,m} \mathcal{R}_{4}(
 \varepsilon_{n},\varepsilon_{m}) (-)^{G_{n}-G_{m}}\langle 
 m|{\bm z}\rangle\{ {\bm \sigma}_1\otimes{\bm\tau}_1\}_{1}\langle{\bm
   z}|n\rangle\cdot\langle n|{\bm\tau}_1|m\rangle,\\ 
\mathcal{H}({\bm z}) & = & N_c
\sum_{\varepsilon_{n}\neq\varepsilon_{v}} \,\frac{1}{
  \varepsilon_{v}-\varepsilon_{n}} \langle
v|{\bm z}\rangle\{ {\bm \sigma}_1\otimes{\bm\tau}_1\}_{0}\langle{\bm
  z}|n\rangle\langle n|\gamma_4|v\rangle\cr 
 &  &  +\; \frac{N_c}{2}\sum_{n,m} \mathcal{R}_{2} ( \varepsilon_{n},
 \varepsilon_{m}) \sqrt{2G_{m}+1}\langle
 m|{\bm z}\rangle\{ {\bm \sigma}_1\otimes{\bm\tau}_1\}_{0}\langle{\bm
   z}|n\rangle\langle n|\gamma_4|m\rangle, \\ 
\mathcal{I}({\bm z}) & = & N_c
\sum_{\varepsilon_{n}\neq\varepsilon_{v}} \,\frac{1}{ \varepsilon_{v}
  - \varepsilon_{n}}(-)^{G_{n}}\langle
v|{\bm z}\rangle {\bm \sigma}_1\langle{\bm z}|n\rangle \cdot\langle 
n|\gamma_4{\bm\tau}_1|v\rangle\cr 
 &  &  +\; \frac{N_c}{2}\sum_{n,m} \mathcal{R}_{2}( \varepsilon_{n},
 \varepsilon_{m}) (-)^{G_{n}-G_{m}} \langle
 m|{\bm z}\rangle {\bm \sigma}_1\langle{\bm z}|n\rangle \cdot \langle 
 n|\gamma_4{\bm\tau}_1|m\rangle,\\ 
\mathcal{J}({\bm z}) & = & N_c \sum_{n^{0}} \,\frac{1}{
  \varepsilon_{v} -\varepsilon_{n^{0}}}\langle
v|{\bm z}\rangle\{ {\bm \sigma}_1\otimes{\bm\tau}_1\}_{0}\langle{\bm
  z}|n^{0}\rangle\langle n^{0}|\gamma_4|v\rangle\cr 
 &  &  +\; N_c \sum_{n,m}\mathcal{R}_{2}( \varepsilon_{n^{0}},
 \varepsilon_{m}) \sqrt{2G_{m}+1}\langle m|{\bm z}\rangle\{
 {\bm \sigma}_1\otimes{\bm\tau}_1\}_{0}\langle{\bm
   z}|n^{0}\rangle\langle  n^{0}|\gamma_4|m\rangle,
\end{eqnarray}
where we have used a standard notation for the irreducible tensor
algebra~\cite{Quantum_angular_theory}.  The proper-time regularization
functions are defined as  
\begin{eqnarray}
\mathcal{R}_{1}(\varepsilon_{n}) & = & -\frac{1}{2\sqrt{\pi}}\,
\varepsilon_{n}\ \int_{1/\Lambda^{2}}^{\infty}\,\frac{du\,}{\sqrt{u}}
\, e^{-u\varepsilon_{n}^{2}},\\
\mathcal{R}_{2}(\varepsilon_{n},\varepsilon_{m}) & = &
\int_{1/\Lambda^{2}}^{\infty}\, du\frac{1}{2\sqrt{\pi
    u}}\frac{\varepsilon_{m}e^{-u\varepsilon_{m}^{2}}-\varepsilon_{n}
  e^{-u\varepsilon_{n}^{2}}}{\varepsilon_{n}-\varepsilon_{m}},\\  
\mathcal{R}_{4}(\varepsilon_{n},\varepsilon_{m}) & = & \frac{1}{2\pi}
\int_{1/\Lambda^{2}}^{\infty}\, du\int_{0}^{1}d\alpha\,
e^{-\varepsilon_{n}^{2}u(1-\alpha)-\alpha\varepsilon_{m}^{2}u}\,\,
\frac{\varepsilon_{n}(1-\alpha) -
  \alpha\varepsilon_{m}}{\sqrt{\alpha(1-\alpha)}},\\   
\mathcal{R}_{5}(\varepsilon_{n},\varepsilon_{m}) & = & 
\frac{1}{2}\frac{\mathrm{sign}\varepsilon_{n}-\mathrm{sign}
  \varepsilon_{m}}{\varepsilon_{n}-\varepsilon_{m}}.
\end{eqnarray}
The $|v\rangle$ denotes the valence quark state and $| n\rangle$ are
the quark eigen-states of the $\chi$QSM Hamiltonian $H(U)$.
$\varepsilon_v$ and $\varepsilon_n$ are the corresponding
eigen-energies, respectively.  $|n^0\rangle$ and $\varepsilon_{n^0}$
are the eigen-states and eigen-energies of the Hamiltonian for the
vacuum $H(U=1)$ (See, for example, Ref.~\cite{WakamatsuBasis}). 

\section{Baryon matrix elements\label{app:bame}}
Eq.~(\ref{eq:axial-model}) can be expressed in the following way:
\begin{eqnarray}
G_{A}^{a}(0) & = & a_{1}D_{a3}^{(8)}+a_{2}d_{pq3}D_{ap}^{(8)} J_{q} +
\frac{a_{3}}{\sqrt{3}}\, D_{a8}^{(8)}J_{3} 
\cr
 & + & \frac{a_{4}}{\sqrt{3}}\, d_{pq3}D_{ap}^{(8)}D_{8q}^{(8)} +
 a_{5}\Big[D_{a3}^{(8)} D_{88}^{(8)} +
 D_{a8}^{(8)}D_{83}^{(8)}\Big] + a_{6}\Big[D_{a3}^{(8)}
 D_{88}^{(8)} - D_{a8}^{(8)}D_{83}^{(8)}\Big],
\label{eq:as}
\end{eqnarray} 
where the numerical results for the axial-vector parameters $a_{i}$
with $M=420\mathrm{MeV}$ are listed in Table~\ref{tab:ai}. 
\begin{table}[h]
\label{tab:ai}
  \centering
  \caption{Numerical values for the axial-vector parameters $a_i$ with
    $M=420$ MeV.}
\begin{tabular}{cccccc}
\hline
$a_{1}$&
$a_{2}$&
$a_{3}$&
$a_{4}$&
$a_{5}$&
$a_{6}$
\tabularnewline\hline
$-3.70$&
$2.50$&
$0.90$&
$-0.18$&
$0.02$&
$0.04$\tabularnewline
\hline
\end{tabular}
\end{table}
The axial-vector parameter $a_{1}$ contains not only the leading
order in rotation and $m_{\mathrm{s}}$ but also a part of the rotational as
well as the $m_{\mathrm{s}}$ operator corrections.  Without
$m_{\mathrm{s}}$ corrections, we have $a_{1}=-3.64$.  All the
axial-vector constants presented in this work can be also reproduced
by these values.  We have the following expressions from
Refs.~\cite{Diakonov:1997mm, semileptonic_SU(3)_CQSM}, respectively:  
\begin{eqnarray}
\frac{F}{D} & = & \frac{5}{9}\cdot\frac{G_{0}+\frac{1}{2}G_{1} +
  \frac{1}{2}G_{2}}{G_{0} + \frac{1}{2}G_{1}-\frac{1}{6}G_{2}},\;\;\;
\frac{F}{D} = \frac{5}{9}\cdot\frac{-a_{1} + \frac{1}{2}a_{2} +
  \frac{1}{2}a_{3}}{-a_{1} + \frac{1}{2}a_{2}-\frac{1}{6}a_{3}},
\end{eqnarray}
and from Ref.~\cite{mixings:2004}
\begin{equation}
G_{i}=ga_{i+1},\;\;\; G_{2}=\frac{2m_{N}}{3f_{\pi}}g_{A}^{0}=ga_{3}
\end{equation}
with $g=\frac{1}{a_{3}}\frac{2m_{N}}{3f_{\pi}}g_{A}^{0}$.  In the
$\chi$QSM, we obtain $g_{A}^{0}=0.367$~\cite{SilvaKUG:2005}  
which yields $g = 2.74$ and the following results:
\begin{equation}
G_{0}^{\chi\mathrm{QSM}}=9.98,\;\;\; 
G_{1}^{\chi\mathrm{QSM}}=6.86,\;\;\; 
G_{2}^{\chi\mathrm{QSM}}=2.47.
\end{equation}
The baryonic transition matrix elements for the collective operators
given in Eq.(\ref{eq:as}) are listed in
Tables~\ref{tab:mat1}--\ref{tab:mat3}. We use the octet-basis
according to Ref.~\cite{CHP}, which means that there is a factor of 
$\sqrt{2}$ involved in calculating the strong coupling constants
from the axial-vector constants for off-diagonal transitions.
\begin{table}[h]
  \centering

\caption{The transition matrix elements between the baryon
  anti-decuplet and the octet for the operators $D_{a3}^{(8)}$ and
  $D_{a8}^{(8)}J_3$.  Those for the operator $d_{pq3} D_{ap} J_q$ are
  simply the same as those for $D_{a3}^{(8)}$, correspondingly.} 

\label{tab:mat1}
\begin{tabular}{c|ccccccc}
\hline
$ $&
$\Theta^{+}n$&
$p_{\overline{10}}n$&
$p_{\overline{10}}\Lambda$&
$p_{\overline{10}}p$&
$\Xi_{\overline{10}}^{+}\Xi^{0}$&
$\Xi_{\overline{10}}^{--}\Sigma^{-}$&
$p_{\overline{10}}\Delta^{+}$\tabularnewline
\hline
$D_{a3}^{(8)}$&
$-\frac{1}{2}\sqrt{\frac{1}{15}}$&
$-\frac{1}{6}\sqrt{\frac{1}{5}}$&
$\frac{1}{2}\sqrt{\frac{1}{30}}$&
$\frac{1}{2}\sqrt{\frac{1}{15}}$&
$\frac{1}{2}\sqrt{\frac{1}{15}}$&
$\frac{1}{2}\sqrt{\frac{1}{15}}$&
$0$
\tabularnewline
$D_{a8}^{(8)}J_3$&
$-\frac{1}{4}\sqrt{\frac{1}{5}}$&
$-\frac{1}{4}\sqrt{\frac{1}{15}}$&
$\frac{1}{4}\sqrt{\frac{1}{10}}$&
$\frac{1}{4}\sqrt{\frac{1}{5}}$&
$\frac{1}{4}\sqrt{\frac{1}{5}}$&
$\frac{1}{4}\sqrt{\frac{1}{5}}$&
$0$
\tabularnewline
\hline
\end{tabular}
\end{table}

\begin{table}[h]
  \centering

\caption{The transition matrix elements between the baryon
  anti-decuplet and the octet for the operators $D_{88}^{(8)} D_{a3}^{(8)}$,
  $D_{83}^{(8)} D_{a8}^{(8)}$, and $d_{pq3}D_{8p}^{(8)} D_{aq}^{(8)}$.} 

\label{tab:mat2}
\begin{tabular}{c|ccccccc}
\hline
$ $&
$\Theta^{+}n$&
$p_{\overline{10}}n$&
$p_{\overline{10}}\Lambda$&
$p_{\overline{10}}p$&
$\Xi_{\overline{10}}^{+}\Xi^{0}$&
$\Xi_{\overline{10}}^{--}\Sigma^{-}$&
$p_{\overline{10}}\Delta^{+}$\tabularnewline
\hline
$D_{88}^{(8)}D_{a3}^{(8)}$&
$-\frac{1}{8}\sqrt{\frac{1}{15}}$&
$-\frac{5}{36}\sqrt{\frac{1}{5}}$&
$-\frac{1}{8}\sqrt{\frac{1}{30}}$&
$0$&
$-\frac{1}{24}\sqrt{\frac{1}{15}}$&
$-\frac{1}{12}\sqrt{\frac{1}{15}}$&
$\frac{5}{144}$\tabularnewline
$D_{83}^{(8)}D_{a8}^{(8)}$&
$\frac{1}{8}\sqrt{\frac{1}{15}}$&
$\frac{1}{36}\sqrt{\frac{1}{5}}$&
$\frac{1}{8}\sqrt{\frac{1}{30}}$&
$0$&
$\frac{5}{24}\sqrt{\frac{1}{15}}$&
$-\frac{1}{12}\sqrt{\frac{1}{15}}$&
$-\frac{1}{144}$\tabularnewline
$D_{8c}^{(8)}D_{a b}^{(8)}d_{cb3}$&
$\frac{1}{12}\sqrt{\frac{1}{5}}$&
$-\frac{1}{18}\sqrt{\frac{1}{15}}$&
$-\frac{1}{12}\sqrt{\frac{1}{10}}$&
$-\frac{1}{12}\sqrt{\frac{1}{5}}$&
$\frac{2}{18}\sqrt{\frac{1}{5}}$&
$-\frac{1}{36}\sqrt{\frac{1}{15}}$&
$-\frac{1}{36}\sqrt{\frac{1}{3}}$\tabularnewline
\hline
\end{tabular}
\end{table}

\begin{table}[h]
  \centering

\caption{The transition matrix elements for the wave-function
  corrections with the baryons states defined in
  Eqs.(\ref{B8},\ref{B10},\ref{wavefunctions}).}  

\label{tab:mat3}
\begin{tabular}{c|cccccc}
\hline
$D_{a3}^{(8)}$&
$\Theta^{+}n$&
$p_{\overline{10}}n$&
$p_{\overline{10}}\Lambda$&
$p_{\overline{10}}p$&
$\Xi_{\overline{10}}^{+}\Xi^{0}$&
$\Xi_{\overline{10}}^{--}\Sigma^{-}$\tabularnewline
\hline
$c_{\overline{10}}^{B}$&
$-\frac{1}{8}\sqrt{\frac{1}{3}}$&
$-\frac{1}{24}$&
$0$&
$-\frac{1}{8}\sqrt{\frac{1}{3}}$&
$0$&
$-\frac{1}{8}\sqrt{\frac{1}{3}}$\tabularnewline
$c_{27}^{B}$&
$\frac{7}{24}\sqrt{\frac{1}{10}}$&
$-\frac{49}{72}\sqrt{\frac{1}{30}}$&
$-\frac{7}{12}\sqrt{\frac{1}{30}}$&
$-\frac{7}{8}\sqrt{\frac{1}{90}}$&
$\frac{7}{36}\sqrt{\frac{1}{10}}$&
$\frac{7}{48}\sqrt{\frac{1}{15}}$\tabularnewline
$d_{8}^{B}$&
$0$&
$-\frac{7}{30}$&
$-\frac{2}{5}\sqrt{\frac{1}{6}}$&
$-\frac{1}{10}\sqrt{\frac{1}{3}}$&
$0$&
$0$\tabularnewline
$d_{27}^{B}$&
$0$&
$-\frac{1}{45}\sqrt{\frac{1}{6}}$&
$\frac{1}{30}$&
$-\frac{1}{15}\sqrt{\frac{1}{2}}$&
$\frac{1}{9}\sqrt{\frac{1}{10}}$&
$-\frac{1}{9}\sqrt{\frac{1}{10}}$\tabularnewline
\hline
\end{tabular}
\vspace{0.3cm}

\begin{tabular}{c|cccccc}
\hline
$D_{a8}^{(8)}J_{3}$&
$\Theta^{+}n$&
$p_{\overline{10}}n$&
$p_{\overline{10}}\Lambda$&
$p_{\overline{10}}p$&
$\Xi_{\overline{10}}^{+}\Xi^{0}$&
$\Xi_{\overline{10}}^{--}\Sigma^{-}$\tabularnewline
\hline
$c_{\overline{10}}^{B}$&
$\frac{1}{16}$&
$\frac{1}{16}\sqrt{\frac{1}{3}}$&
$0$&
$\frac{1}{16}$&
$0$&
$\frac{1}{16}$\tabularnewline
$c_{27}^{B}$&
$-\frac{3}{16}\sqrt{\frac{1}{30}}$&
$\frac{7}{48}\sqrt{\frac{1}{10}}$&
$\frac{1}{8}\sqrt{\frac{1}{10}}$&
$\frac{3}{16}\sqrt{\frac{1}{30}}$&
$-\frac{1}{8}\sqrt{\frac{1}{30}}$&
$-\frac{1}{36}\sqrt{\frac{1}{5}}$\tabularnewline
$d_{8}^{B}$&
$0$&
$\frac{1}{20}\sqrt{\frac{1}{3}}$&
$\frac{1}{10}\sqrt{\frac{1}{2}}$&
$\frac{3}{20}$&
$0$&
$0$\tabularnewline
$d_{27}^{B}$&
$0$&
$\frac{1}{30}\sqrt{\frac{1}{2}}$&
$-\frac{3}{20}\sqrt{\frac{1}{3}}$&
$\frac{3}{10}\sqrt{\frac{1}{6}}$&
$-\frac{1}{2}\sqrt{\frac{1}{30}}$&
$\frac{1}{2}\sqrt{\frac{1}{30}}$\tabularnewline
\hline
\end{tabular}
\vspace{0.3cm}

\begin{tabular}{c|cccccc}
\hline
$d_{cb3}D_{a b}^{(8)}J_{c}$&
$\Theta^{+}n$&
$p_{\overline{10}}n$&
$p_{\overline{10}}\Lambda$&
$p_{\overline{10}}p$&
$\Xi_{\overline{10}}^{+}\Xi^{0}$&
$\Xi_{\overline{10}}^{--}\Sigma^{-}$\tabularnewline
\hline
$c_{\overline{10}}^{B}$&
$-\frac{5}{16}\sqrt{\frac{1}{3}}$&
$-\frac{5}{48}$&
$0$&
$-\frac{5}{16}\sqrt{\frac{1}{3}}$&
$0$&
$-\frac{5}{16}\sqrt{\frac{1}{3}}$\tabularnewline
$c_{27}^{B}$&
$-\frac{11}{48}\sqrt{\frac{1}{10}}$&
$\frac{77}{144}\sqrt{\frac{1}{30}}$&
$\frac{11}{24}\sqrt{\frac{1}{30}}$&
$\frac{33}{144}\sqrt{\frac{1}{10}}$&
$-\frac{11}{72}\sqrt{\frac{1}{10}}$&
$-\frac{11}{96}\sqrt{\frac{1}{15}}$\tabularnewline
$d_{8}^{B}$&
$0$&
$\frac{7}{60}$&
$\frac{1}{5}\sqrt{\frac{1}{6}}$&
$\frac{1}{20}\sqrt{\frac{1}{3}}$&
$0$&
$0$\tabularnewline
$d_{27}^{B}$&
$0$&
$-\frac{2}{45}\sqrt{\frac{1}{6}}$&
$\frac{1}{15}$&
$-\frac{2}{15}\sqrt{\frac{1}{2}}$&
$\frac{2}{9}\sqrt{\frac{1}{10}}$&
$-\frac{2}{9}\sqrt{\frac{1}{10}}$\tabularnewline
\hline
\end{tabular}
\end{table}

The transition matrix elements of the $p_{\overline{10}}\to\Delta^+$
process for the wave-function corrections are given as
\begin{eqnarray}
\langle\Delta^{+}|\,\frac{1}{2} D_{33}^{(8)}\,|p_{\overline{10}}\rangle & = &
d_{8}^{B}\,\frac{1}{3}\,\sqrt{\frac{1}{5}}\,+d_{27}^{B}\,\frac{1}{9}\,
\sqrt{\frac{1}{30}}\,+a_{27}^{B}\,\frac{5}{9}\,\sqrt{\frac{1}{30}}, \cr 
\langle\Delta^{+}|\,\frac{1}{2} D_{38}^{8}\, J_{3}\,|p_{\overline{10}}\rangle & = & 0,\cr
\langle\Delta^{+}|\, \frac{1}{2} d_{pq3}D_{3q}^{(8)}J_{p}\,|p_{\overline{10}}\rangle & = & 
- d_{8}^{B}\,\frac{1}{6}\sqrt{\frac{1}{5}}+d_{27}^{B}\,\frac{2}{9}
\,\,\sqrt{\frac{1}{30}}-a_{27}^{B}\,\frac{5}{9}\,\sqrt{\frac{1}{30}}.
\nonumber  
\end{eqnarray}

\newpage
\end{appendix}

\end{document}